\documentclass{article}
\usepackage{emulateapj}
\usepackage{epsf}
\usepackage{graphicx}

\def\lea{\mathrel{<\kern-1.0em\lower0.9ex\hbox{$\sim$}}}
\def\gea{\mathrel{>\kern-1.0em\lower0.9ex\hbox{$\sim$}}}

\begin{document}


\title{The Globular Cluster Systems of Five Nearby Spiral Galaxies: New Insights from \emph{Hubble Space Telescope} Imaging}

\author{Rupali  Chandar\altaffilmark{1}, Bradley Whitmore\altaffilmark{1}, and Myung Gyoon  Lee\altaffilmark{2,3}}
\altaffiltext{1}{Space Telescope Science Institute, 3700 San Martin Dr., Baltimore, MD 21218} 
\altaffiltext{2}{Astronomy Program, SEES, Seoul National University, Seoul 151-742, Korea}
\altaffiltext{3}{Visiting Investigator at the Department of Terrestrial Magnetism, Carnegie Institution
of Washington, 5241 Broad Branch Road, N.W., Washington, D.C. 20015}
\affil{email: rupali@stsci.edu, whitmore@stsci.edu, mglee@astrog.snu.ac.kr }

\begin{abstract} 

We use available multifilter {\it Hubble Space Telescope} (\textit{HST})
WFPC2 imaging of five (M81, M83, NGC~6946, M101, and M51, in order of
distance) low inclination, nearby spiral galaxies to study ancient
star cluster populations.  Combining rigorous selection criteria to
reject contaminants (individual stars, background galaxies, and
blends) with optical photometry including the U bandpass,
we unambiguously detect ancient globular cluster (GC) systems in each galaxy.
We present luminosities, colors, and size (effective radius)
measurements for our candidate GCs.  These are used to estimate
specific frequencies, assess whether intrinsic color distributions are
consistent with the presence of both metal-poor and metal-rich GCs,
and to compare relative sizes of ancient clusters between different
galaxy systems.

M81 globulars have intrinsic color distributions which are very
similar to those in the Milky Way and M31, with $\sim40$\% of sample
clusters having colors expected for a metal-rich population.  The GC
system in M51 meanwhile, appears almost exclusively blue and metal
poor.  This lack of metal-rich GCs associated with the M51 bulge
indicates that the bulge formation history of this Sbc galaxy may have
differed significantly from that of our own.  Ancient clusters in M101
and possibly in NGC~6946, two of the three later-type spirals in our
sample, appear to have luminosity distributions which continue to rise
to our detection limit ($M_V\sim-6.0$), well beyond the expected
turnover ($M_V\sim-7.4$) in the luminosity function.  This is
reminiscent of the situation in M33, a Local Group galaxy of similar
Hubble type.  The faint ancient cluster candidates in M101 and
NGC~6946 have properties (colors and $r_{eff}$) similar to their more
luminous counterparts, and we suggest that these are either
intermediate age ($3-9$~Gyr) disk clusters or the low mass end of the
original GC population.  Potentially, these lower mass clusters
weren't destroyed due to different dynamical conditions relative to
those present in earlier-type galaxies.  If the faint, excess GC
candidates are excluded, we find that the specific frequency ($S_N$)
of ancient clusters formed in {\it later-type} spirals is roughly
constant, with $S_N=0.5\pm0.2$.  If we consider only the blue,
metal-poor clusters in the {\it early-type} spiral M81, this galaxy is
also consistent with having formed a ``universal'' specific frequency
of halo GC population, with a value of $S_N\sim0.6$.  By combining the
results of this study with literature values for other systems, we
find that the total GC specific frequencies in spirals appear to
correlate best with Hubble type and bulge/total ratio, rather than
with galaxy luminosity or galaxy mass.
 
\end{abstract}
 

\section{INTRODUCTION}

Old stellar populations, both old stars and star clusters, provide
unique insight into the early assembly history of their parent
galaxies.  For example, in the Milky Way, ages, abundances, and
kinematics of these two stellar populations portray a relatively
quiescent early evolution, with no significant merging since the
formation of the Galactic thick disk $\sim12$ Gyr ago (see e.g., Wyse
2000 and references therein).  Subpopulations of globular clusters
(GCs), as luminous tracers of mass, are found in the halo (metal poor,
little rotation), associated with the bulge (metal rich, centrally
concentrated), and also with the thick disk (metal rich, rotationally
supported) (e.g., Zinn 1985; Armandroff 1989; Minniti 1995; Cote 1999).

Old clusters in Andromeda show both similarities to and differences
with their Galactic counterparts.  The luminosity, metallicity, and
size distributions of GCs in M31 and the Milky Way appear extremely
similar (e.g., Crampton et~al.\  1985; Perrett et~al.\  2002; Barmby,
Holland, \& Huchra 2002).  However, recent kinematic studies suggest
that a ``cold'' rotating thin disk of ancient GCs (covering the entire
range of metallicities) exists in M31, which implies that M31 could
not have undergone any significant accretion events since the
formation of these objects (Morrison et~al.\  2003).  Any theory for the
formation of Andromeda will have to simultaneously explain this result
for the GC system, and the recent discovery of intermediate age
($\sim6-8$ Gyr), metal-rich stars (from main sequence fitting) in the
halo of M31 (Brown et~al.\  2003).  A number of works in different
portions of the M31 halo have established that the metallicity
distribution of the field stars differs substantially from that of the
GC distribution (e.g., Durrell, Harris, \& Pritchet 2001; Reitzel \&
Guhathakurta 2002).  With their additional age constraints, Brown et
al. (2003) suggest that a late merging event is the most likely
scenario for the presence of these halo stars.

M33 is the final and latest-type spiral galaxy in the Local Group.
The early work of Mould \& Kristian (1986) suggested that M33 halo
stars have a very low mean metallicity, $\sim-2$~dex, with a small
spread.  However, more recent analysis of the same field indicates
that this location is still dominated by (metal-rich) M33 disk stars,
although there may be a very small contribution from a (metal-poor)
stellar halo (Tiede, Sarajedini, \& Barker 2004).  Surprisingly,
despite its low luminosity (mass), M33 has a relatively large GC
population, with the majority of these having halo kinematics
(Christian \& Schommer 1988; Schommer et~al.\  1991; Chandar et~al.\  
2002).  The M33 GC system appears quite different from those in
the Galaxy and M31 in at least three ways: {\it i)} there is evidence
from horizontal branch morphology (Sarajedini et~al.\  1998) and
spectroscopic line indices (Chandar et~al.\  2002) that the halo
clusters have a much larger age spread than those found in the Galaxy
and M31 (although see Larsen et~al.\  2002 for a different viewpoint);
{\it ii)} the luminosity function of ancient M33 clusters appears to
continue to rise beyond the $M_V\sim-7.4$ cutoff seen in the Galactic
and M31 GC systems; and {\it iii)} the estimated total population of
$75\pm14$ GCs (Chandar, Bianchi, \& Ford 2001) gives this galaxy a
higher mass normalized GC population ($T=3.8\pm0.7$) than the two
earlier-type spirals ($T=1.3\pm0.2$ and $1.6\pm0.4$ for the Milky Way
and M31 respectively) in the Local Group.

As one moves beyond the Local Group, it becomes much more difficult to
access individual stars, and GCs become the ancient stellar population
tracer of choice.  However, ground based studies of late-type galaxies
beyond the Local Group, even of galaxies at high inclination, have
shown mixed results.  Contamination by foreground stars and background
galaxies can be a major problem.  For example, using follow-up
spectroscopy, Beasley \& Sharples (2000) confirmed only 14/64 and 1/55
GC candidates in NGC~253 and NGC~55 respectively.

The depth and resolution of imaging possible with the {\it Hubble
Space Telescope} (\textit{HST}) has transformed the field of extragalactic GC
research over the last decade, and motivated significant progress in
understanding the formation and evolution of GC systems, particularly
in elliptical and lenticular galaxies.  One very interesting result is
widespread evidence for bimodal color distributions in early-type
galaxy GC systems over a large range of luminosity, indicative of
multiple episodes of cluster formation even in lower mass ellipticals
(e.g., Kundu \& Whitmore 2001; Gebhardt \& Kissler-Patig 1999).
Despite evidence that GCs exist in all massive galaxies, as well as in
a number of lower mass systems, our understanding of the formation of
spiral galaxy GC systems remains much poorer than for early-type
galaxies.  

A recent \textit{HST} study of seven edge-on spirals (out to the distance of
Virgo) has made important progess in this direction.  Goudfrooij et
al. (2003) studied the GC systems of seven edge-on
spirals, from Sa to Sc, using V and I band \textit{HST} WFPC2 imaging.  They
find that the specific frequency ($S_N$) of GCs in spirals with Hubble
types later than Sb are all consistent with a value of $0.55\pm0.25$,
supporting the concept of a ``universal'' old halo population
in later-type spirals.  
Because a few earlier-type spirals are known to have larger specific
frequencies than this value, Goudfrooij et~al.\  (2003) suggest that a
second, metal rich ``bulge'' population in galaxies with large
bulge/total (B/T) ratios could explain current observations of spiral
GC systems.  This fits into the Forbes, Brodie, \& Larsen (2001)
scenario, where a ``universal'' metal-rich GC population forms in
association with both spiral bulges and elliptical spheroids.
One goal of this paper is to look for further evidence
supporting or dismissing these concepts of ``universal'' halo and
bulge GC systems.

In order to expand the number of spirals which have detailed GC system
information, additional samples of these ancient objects are needed
which can be followed up with ground based spectroscopy (to measure
ages, abundances, and velocities).  Due to the need for eventual
spectroscopy, in this work we study spirals within 10 Mpc.
Furthermore, because it appears that GCs may sometimes reside in thin
disks, we restrict target selection to include galaxies with
relatively low inclinations ($\lea 65^{\circ}$).  One potential
difficulty in studying ancient star clusters in late-type galaxies, is
the fact that these systems usually have on-going cluster formation in
the disk.  In terms of numbers, clusters with ages younger than
several Gyr often completely overwhelm their older counterparts.  For
example, in M33 there are currently $\sim50$ known ancient GCs, but
several hundred known younger clusters (e.g., Christian \& Schommer
1988; Chandar, Bianchi, \& Ford 1999a, 2001).  With access only to
optical photometry, there are degeneracies in broadband colors among
age, reddening, and metallicity, which can lead to significant
contamination of an old cluster sample by reddened young clusters.
However, with an appropriate filter combination,
these degeneracies can be sorted out. In particular, the U bandpass in
combination with redder filters provides crucial information to
differentiate among ancient and (reddened) young objects.

In this work we attempt to broadly characterize GC systems in five
nearby spirals: M81, M83, NGC~6946, M101, and M51.  These target
galaxies were chosen because they are nearby,
and they have multifilter HST WFPC2 imaging observations available; in
particular there is at least some U band information.
We are interested in fundamental parameters, such as the total number
of GCs in each galaxy, their specific frequencies, the luminosity and
color distributions, and finally the size distribution of GCs.
Global properties of the target galaxies are given in Table 1.  This
paper is organized as follows: $\S2$ gives background information
regarding the current status of our knowledge of the GC systems in the
target galaxies.  We explicitly describe the advantages of this work
over previous studies; $\S3$ describes the data and reduction; $\S4$
presents luminosity, color, and size distributions, as well as GC
specific frequencies; and $\S5$ discusses the global properties of the
GC systems, and their consistency within the framework of
``universal'' ancient cluster subsystems.  In $\S6$ we summarize the
main results of this work.

\begin{deluxetable}{lllcrcrrrl}             
\tablewidth{0pc}             
\tablecaption{GLOBAL PROPERTIES OF SAMPLE GALAXIES}            
\small
\tablehead{             
\colhead{Galaxy} & \colhead{RA(J2000)}  & 
\colhead{DEC (J2000)} & \colhead{Type (RSA)\tablenotemark{a}}  & 
\colhead{$A_V$\tablenotemark{b}} &
\colhead{$m-M$\tablenotemark{c}} 
}             
\startdata           
M81 & 08:23:56 & +71:01:45 & Sab (2) & 0.266 & $27.8\pm0.2$    \\
M83 & 13:37:01 & $-$29:51:57 & \phn Sc (5) & 0.218  & $28.25\pm0.15$   \\
NGC6946 & 20:34:52 & +60:09:14 & Scd (6) & 1.133 & $28.85\pm0.15$    \\
M101 & 14:03:12 & +54:20:55 & Scd (6) & 0.028 & $29.21\pm0.17$    \\
M51 & 13:29:53 & +47:11:43 & Sbc (4) & 0.115 & $29.62\pm0.15$   \\
\tablenotetext{a}{From de Vaucouleurs et~al.\  (1991)}
\tablenotetext{b}{Foreground extinction values are from Schlegel et~al.\  (1998)}
\tablenotetext{c}{Galaxy distances are taken from the following sources: M81 -- Freedman et~al.\  1994; M83 -- Thim et~al.\  2003 ; NGC~6946 -- Karachentsev, Sharina, \& Huchtmeier 2000; M101 -- Stetson et~al.\  1998; M51 -- Feldmeier et~al.\  1997 }
\normalsize
\enddata             
\end{deluxetable}

\section{Past Results on Selected Galaxies}

{\it M81:} To date, the globular cluster population in M81 has been
studied in a handful of works.  Using BVR colors and magnitudes,
Perelmuter \& Racine (1995) found an excess of $\sim70$ objects within
11 kpc of the center of M81.  After completeness corrections, they
estimated the total GC population of this early-type spiral to be
$210\pm30$.  Followup spectroscopy of cluster candidates chosen from
color and magnitude cuts (plus proper motion information)
confirmed 25 of these objects to be bona fide GCs (another 19 are
listed as probable GCs and 29 were found to be either background
galaxies or foreground stars; Perelmuter, Brodie, \& Huchra 1995).
The mean derived metallicity for the GCs in the Perelmuter et
al. (1995) study is [Fe/H]$=-1.48\pm0.19$. 
Schroder et~al.\  (2001) have recently obtained individual metallicity
measurements for 16 GC candidates in M81 (with target selection from
the Perelmuter works).  Fifteen of these have spectra consistent with
bona fide globulars.  They find evidence from a sample of 44 total GCs
that red (metal-rich) objects rotate in the same sense as the gas in
the M81 disk, and that the blue (metal-poor) clusters have halo-like
kinematics, with little evidence for rotation.

We previously studied the cluster system in M81 using BVI \textit{HST} WFPC2
imaging (Chandar, Ford, \& Tsvetanov 2001; Chandar, Tsvetanov, \& Ford
2001).  We found that in addition to an ancient GC system, M81
(despite being an early type spiral) has formed compact young
clusters, although these tend to be lower in mass than older GCs.
Here, we re-analyse the eight \textit{HST} fields used in our previous study,
and add three more which are now available.  However the biggest
advantage of this work over our previous effort is the inclusion of
available U band observations, allowing us to make a more detailed
study of the M81 GC system (the focus of our previous work was on the
young cluster properties).

{\it M101:} Bresolin et~al.\  (1996) studied \textit{HST} WFPC2 imaging of a
single field near the center of M101, and detected 41 compact
clusters.  Most of these have colors which are too blue to be ancient
GCs.  There are however, five clusters which have (B-V) and (V-I)
colors consistent with those of ancient GCs in the Milky Way.  Because
Bresolin et~al.\  (1996) have published $(U-B)$ photometry for only one
of these objects, it is unclear whether the others are reddened young
clusters, or really ancient cluster candidates.
In this work, we revisit the field studied by Bresolin et
al. (1996), but create a deep, drizzled image 
from all available observations.  Our deep image of a central field
pointing in M101 reveals over 400 compact but resolved clusters.
Properties of the entire cluster population will be presented in a
separate work (Chandar et~al.\  2004, in prep.).  Here we include U band
photometry to confirm the existence of a GC system in this
late-type spiral.

{\it M51:} There have been several recent studies of the cluster
system in M51 (e.g., Bik et~al.\  2003; Lamers et~al.\  2002; Larsen
2000).  However, these have concentrated primarily on the large number
of young (massive) clusters, with little mention of the ancient
cluster system in this Milky Way-like galaxy (Sbc).

{\it M83 and NGC~6946:}
To date, there has been little  published on the ancient
cluster systems of these galaxies.  Larsen (2002) noted the existence of
three clusters with colors consistent with those of GCs in a single
WFPC2 pointing in NGC~6946.

\section{DATA REDUCTION, CLUSTER SELECTION, AND PHOTOMETRY}

\subsection{Data and Reduction}

\begin{figure}
\plotone{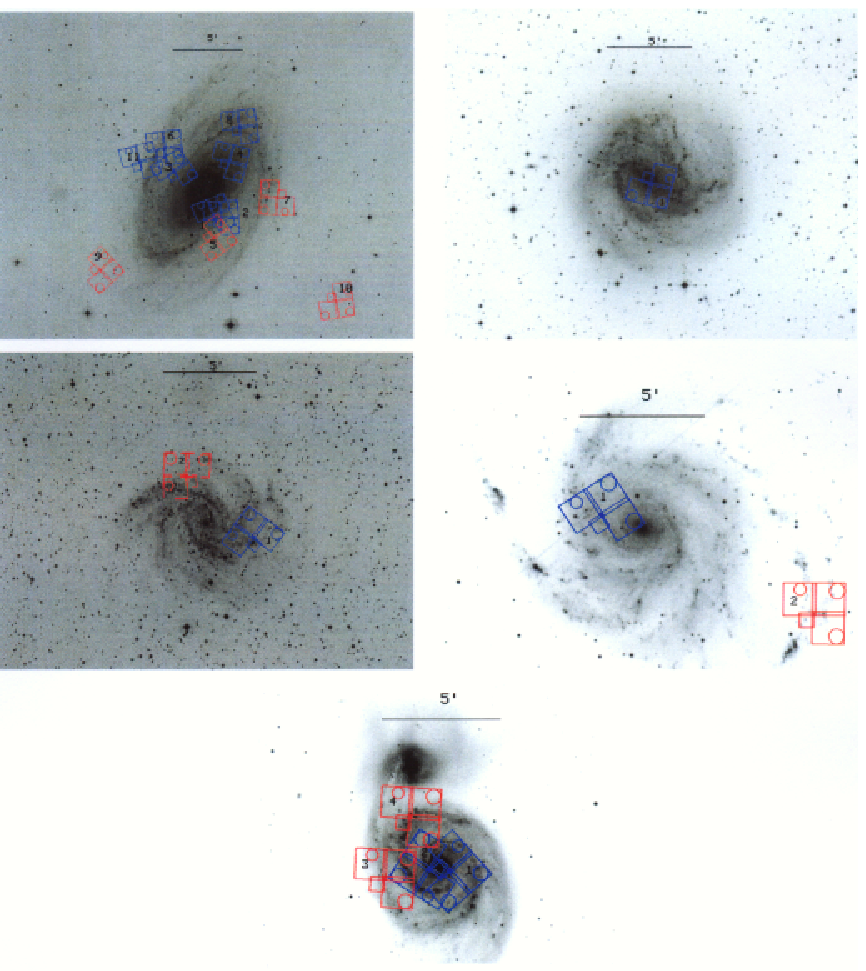}
\caption{The \textit{HST} WFPC2 field pointings used in this study are
overlaid on Digitized Sky Survey images of each target galaxy (M81,
M83, NGC~6946, M101, and M51).  A scale of five arcminutes is shown in
each figure, and WFPC2 footprints shown in blue include U band
imaging.
\label{fov}}
\end{figure}

Available \textit{HST} WFPC2 observations for each galaxy were downloaded
from the archive using the ``on-the-fly'' calibration system, which
automatically uses the best reference files for calibration.  The
WFPC2 pipeline steps include: bad pixel masking, A/D correction, bias
and dark subtraction, and flat field correction.  The locations of the
fields are shown in Figure~\ref{fov} for each target galaxy.  Because we rely
on what is available (taken for a host of different projects with a
variety of filters, exposure times, etc.), we first briefly summarize
basic information for each target galaxy, and then give a general
recipe for reduction.  Information for the fields used in this work,
such as the proposal identification, specific filters and exposure
times are compiled in Table 2.

M81 \textit{HST} WFPC2 observations include seven fields imaged in $UBVI$,
three fields imaged in $BVI$, and an 11th outer field with $VI$
imaging (see Table 2 for details).  We include this outer field
because its large projected distance ($\sim12.5$ arcmin) 
along the semi-minor axis
makes it unlikely that reddened young clusters reside here,
and its location provides a glimpse further out into the halo of this
bulge-dominated galaxy than other available multi-filter \textit{HST} fields.
While \textit{HST} imaging does not provide large coverage in M81, it does allow
us to probe deeper than previous ground-based surveys.

For M83, we use a single \textit{HST} WFPC2 field pointing taken in $UVI$
filters.  While this does not provide much
coverage beyond the central portions in this galaxy, evidence for an
ancient cluster system would be interesting and potentially important
for followup observations.  NGC6946 has one WFPC2 field imaged in
$UBVI$ and a second imaged in $BVI$.

The observations used for M101 were taken between 1994 and 2000 with
the WFPC2, and will be described in detail, along with basic data
reduction and cluster selection, in an upcoming paper.  Here we
provide a brief summary.  Field 1 was observed in UBVI bands, and
field 2 in BVI.  Both pointings were taken for the Cepheid Key
Distance Project, and hence had enough observations with small, random
offsets that we were able to drizzle these images in order to recover
resolution from the undersampled WF CCDs.

Observations of M51 and its nearby companion (NGC~5195) were taken for
a variety of projects. Field 1 is imaged in UBVI, Field 2 in BVRI with
some overlapping U band, and fields 3 and 4 in BVI.  NGC~5195 is covered
in VI filters; results from this field are discussed in Lee, Chandar, \&
Whitmore (2004, in prep).
Taken together, the five fields cover $\sim50$\% of the two-galaxy
system.  Details of the reduction will be presented in our upcoming
paper, which focuses on the age distribution and other properties of
the numerous young massive clusters detected in this interacting
system.

In general, for each field and filter combination, available
observations were combined in pairs to eliminate cosmic rays, after
first checking the alignment.
Combined images were corrected for geometric distortion, as described
in Holtzman et~al.\  (1995).  For field 7 in M81 the three stepped
observations in each of the BVI filters were shifted and combined.

\begin{deluxetable}{lcccccc}
\tablecaption{SUMMARY OF HST WFPC2 FIELD OBSERVATIONS}
\tablehead{
\colhead{FIELD}    & 
\colhead{proposid} &
\multicolumn{5}{c}{Filters and Total Exposure Times [sec]}
\\
\colhead{}   & 
\colhead{}  & 
\colhead{U} & 
\colhead{B} &
\colhead{V} & 
\colhead{R} & 
\colhead{I}
}
\startdata
M81--1 & 6139 & F336W, 1160  & F439W, 1200  & F555W, 900 & F675W, 900 & F814W, 900  \\
M81--2 & 5480 & F336W, 1200 & F439W, 600 & F555W, 300 & F675W, 300 & F814W, 300 \\
M81--3 & 7909 & F300W, 3200 & F450W, 3100 & F606W, 800 & \nodata & F814W, 800 \\
M81--4 & 7909 & F300W, 7300 & F450W, 4300 & F606W, 2000 & \nodata & F814W, 2200 \\
M81--5 & 9073 & \nodata & F450W, 2000 & F555W, 2000 & \nodata & F814W, 2000 \\ 
M81--6 & 5397 & F336W, 1800 & F439W, 1200 & F555W, 800 & \nodata & F814W, 800 \\
M81--7 & 7351 & \nodata & F439W, 2200   & F555W, 1300  & \nodata & F814W, 1300   \\
M81--8 & 5397 & F336W, 1800 & F439W, 1200 & F555W, 800 & \nodata & F814W, 800 \\
M81--9 & 9634 & \nodata & F450W, 2000 & F606W, 1000 & \nodata & F814W, 800 \\
M81--10 & 9086 & \nodata & \nodata & F606W, 5200 & \nodata & F814W, 5500 \\
M81--11 & 8061 & F300W, 1500 & F450W, 5800 & F606W, 8700 & \nodata & F814W, 3000 \\
    &  &  &   &   &   &  \\
M83--1 & 8238 & F300W, 2100 & \nodata & F547M, 930 & \nodata & F814W, 710 \\ 
    &  &  &   &   &   &  \\
NGC~6946--1 & 8715 & F336W, 3000 & F439W, 2200 & F555W, 600 & \nodata & F814W, 1400 \\
NGC~6946--2 & 9073 & \nodata & F450W, 2000 & F555W, 2000 & \nodata & F814W, 2000 \\ 
    &  &  &   &   &   &  \\
M101--1 & 5397 & F336W, 1200 & F439W, 1100 & F555W, 13200 & \nodata & F814W, 4600 \\
M101--2 & 5397 & \nodata & F439W, 1050 & F555W, 4200 & \nodata & F814W, 4800 \\ 
    &  &  &   &   &   &  \\
M51--1 & 7375 & F336W, 1200 & F439W, 1100 & F555W, 1200 & \nodata  & F814W, 1000 \\
M51--2 & 5777 & F336, 1200\tablenotemark{a} & F439W, 1400 & F555W, 600 & F675W, 600 & F814W, 600 \\
M51--3 & 9073  & \nodata  & F450W, 2000 & F555W, 2000  & \nodata & F814W, 2000 \\
M51--4 & 9073  & \nodata  & F450W, 2000 & F555W, 2000 & \nodata  & F814W, 2000 \\ 
\tablenotetext{a}{The U band observations for M51-2 were taken at a somewhat different orientation and pointing.  The overlap region is approximately one Wide Field CCD}
\normalsize
\enddata
\end{deluxetable}

\subsection{Object Detection, Star Cluster Selection, and Photometry}

\subsubsection{Detection}

To identify star clusters, we use morphological information provided
by SEXTRACTOR (Bertin \& Arnouts 1996), to separate true clusters from
contaminants such as individual stars, background galaxies, and
blends.  SEXTRACTOR performed well in our moderately crowded stellar
fields.  Detection was run on the V band images (either drizzled or
combined) for each field, since in all cases these were the deepest
and/or had the best resolution.  We used a threshold of $4\sigma$
above the local background level, in order to avoid large numbers of
detections of very faint objects in our variable, moderately crowded
fields.  In addition to the output from SEXTRACTOR, point spread
function (PSF) fitting was performed on each object, using the IRAF
task ALLSTAR (Stetson 1987).  The PSF was created by automatically
choosing bright, isolated stars using size, shape, and neighbor
information.

\subsubsection{Cluster Selection}

Cluster candidates were selected to be more extended than the PSF and
have low ellipticity values.  This eliminated the majority of
individual stars, background galaxies, and blends.  The primary
remaining source of contamination in our cluster catalogs is from
blends of a few superposed stars (although a number of these were
eliminated from the ellipticity cut).  Finally, each object was
visually inspected, and blends (which are defined as objects which
have large scatter in the central portions relative to the best fit
Moffat profile) were eliminated.  This pipeline provided final
star cluster catalogs in each galaxy.  Independent checks by BW and RC
in M101 showed that very few ($\sim8$ out of $\sim400$) extended
objects which appear to be star clusters were missed by this
algorithm, particularly at the brighter end.  We therefore make the
assumption that we are missing a small percentage of clusters to
$V\sim23.0$ in M101, and that our algorithm is similarly successful
for the other sample galaxies down to comparable completeness limits
(discussed in $\S3.5.2$).

\subsubsection{Photometry}

Because how extended clusters appear in \textit{HST} images depends on both
their intrinsic size and galaxy distance,
we used somewhat different techniques to select clusters in M81
(the closest galaxy in our sample) from those used on M83, NGC~6946, M101,
and M51.  For M81, photometry (using the PHOT task in IRAF) was
performed on clusters using a 10 pixel radius aperture.  For the more
distant galaxies, we used a 3 pixel (non-drizzled) radius, in order to
minimize the contamination from nearby objects.  While this technique
provides robust cluster colors (which are negligibly affected by
aperture corrections; Holtzman et~al.\  1995), there is a
significant fraction of light outside this radius, which must be
corrected for when studying the distribution of total cluster
luminosities.

Here, we describe our general technique for deriving approximate
aperture corrections, by using M51 as a (representative) example.  In
order to measure aperture corrections from 3 to 5 pixels (hereafter
$\Delta m_{3->5}$) for our extended sources, we identified relatively
isolated clusters on the PC CCD and WF CCDs.  In general, these
objects were typically young star clusters, since young massive clusters tend
to be more numerous than ancient clusters in later type spirals.
An average, empirical aperture correction was then obtained by
measuring the mean magnitude differences in 5 and 3 pixel aperture
radii.  For M51, we find mean $\Delta m_{3->5}$ values of $-0.290$ and
$-0.245$ for the PC and WF CCDs respectively.  These values were
compared with table 1 of Larsen \& Brodie (2000), where they have
tabulated aperture corrections based on synthetic King model profiles.
We find that our values are slightly smaller than their KING30
profiles (concentration parameter 30), for a synthetic cluster with a
typical FWHM value of 1.0 pixels.  Overall, we find that M51 clusters
have a FWHM of $\sim0.8$ pixels (Lee, Chandar, \& Whitmore, 2004).
(The description and results of cluster size measurements is presented
in $\S4.4$.)  This gives further confidence in the empirically derived
aperture corrections, since our smaller intrinsic cluster size would
result in a smaller simulated aperture correction, bringing the two
values into excellent agreement.  Because very few of the clusters in
our M51 GC sample are isolated within a 30 pixel radius, we use table 1
in Larsen \& Brodie (2000) to complete the aperture correction to an
infinite radius for the WF CCDs, and use 2 isolated sources in our PC
images.  The final aperture corrections in the PC and WF CCDs for M51
are $-0.39$ and $-0.31$ respectively.  However, we note that aperture
corrections are strongly dependent on intrinsic object size, and for
clusters with sizes as large as 2.0 pixels ($R_{eff}=12~\mbox{pc}$)
the total V magnitude error will be $\sim0.7$ magnitudes.  Hence it
should be kept in mind that our aperture corrections are for a
typically sized cluster; individual clusters will vary.  We note that
this will have a negligible effect on our color estimates, since the
same size aperture is used for each bandpass.

The following steps were used to transform measured broadband WFPC2
instrumental magnitudes to Johnson-Cousins $U$, $B$, $V$, $R$ and $I$
magnitudes: {\it (i)} the instrumental magnitudes were corrected for
the charge-transfer efficiency (CTE) loss, using the prescription
given by A. Dolphin (2000; see
http://www.noao.edu/staff/dolphin/wfpc2\_calib/
for updated calibrated
information); {\it (ii)} the corrected instrumental magnitudes were
converted to standard Johnson-Cousins $U$, $B$, $V$, and $I$
magnitudes.  Using Equation 8 and Table 7 of Holtzman et~al.\  (1995),
the magnitudes were derived iteratively using WFPC2 observations in
two filters, with
all zeropoints substituted from Dolphin (2000), except for the F300W
filter (zeropoint for this filter comes from the WFPC2 Handbook).  $U$
band magnitudes are taken from the coupling of the $U$ and $B$
filters, $B$ magnitudes from the $B$ and $V$ filter combination, and
$V$ and $I$ magnitudes from the $V$ and $I$ filter solution.

We made explicit comparison of the photometry for individual objects
presented here with that from previous works in upcoming papers on the
young cluster systems of M51 and M101.  For M51, a photometric
comparison with clusters studied in Larsen (2000) are in good
agreement -- the mean difference in the V band magnitudes is 0.002,
and the mean difference in color $\Delta~(B-V)$ is $0.048$, in the sense
that our $(B-V)$ color is slightly {\it redder} than that given in
Larsen (2000).  For M101, our comparison with the work of Bresolin et
al. (1996) shows larger differences.  The mean difference in both
V magnitude and color of $\Delta~(B-V)$ is $0.066$.

\subsection{Cluster Reddening Distribution}

\subsubsection{Deriving Ages and Reddening for Clusters}

The final step is to separate ancient globular cluster candidates from
the more numerous young massive clusters found in these galaxies (M81
is the exception, with a higher fraction of luminous, ancient clusters
than comparably bright young clusters).  Because morphologically young
and old clusters are indistinguishable, at this point we used colors
to separate them.  However, there remains the ambiguity of separating
truly ancient, red clusters from young, highly reddened objects.  This
task becomes much easier when there are a minimum of three broadband
filters, {\it particularly including the U bandpass}.  Here we briefly
describe using UBVI observations of field 1 in both M51 and M101 to
study the statistics of the $E(B-V)$ distribution of stellar clusters,
which provides information on the contamination of our ancient cluster
sample from reddened young clusters.  The cluster system of NGC~6946
has been studied previously by Larsen (2002).  M83 only has UVI
filters, making the derived extinction distribution less robust than
in M51 and M101.

In order to determine the age and reddening for each cluster, we use a
modification of the technique described in detail in Bik et~al.\  (2003)
(the Bik et~al.\  version of the fitting routine was kindly made
available to us by H. Lamers).  We compared the observed magnitudes
with spectral energy distributions derived from the theoretical
evolutionary synthesis models of Bruzual \& Charlot (2000; hereafter
BC00).  These spectral synthesis models are available for a number of
metallicities; however, due to the well known
age-metallicity-reddening degeneracy in integrated cluster colors, we
initially assumed the solar model for comparison with the M51 and M101
clusters, in order to best match the young cluster population.
Observations of HII regions in these galaxies establish that the
current metallicity of the gas is approximately solar (e.g., Diaz et
al. 1991; Hill et~al.\  1997).  Tests establish that this assumption has
a negligible effect on the derived ages and extinction values for
younger stellar populations ($\leq 1$ Gyr), but preferentially effects
the ages estimated for older clusters, where metallicity influences
become more pronounced than age influences in the integrated colors.
However, since we are only interested in selecting the ancient cluster
populations and not in their precise ages (which have to wait for
integrated spectroscopy), the integrated colors are sufficient to
separate young and old single stellar populations.

Details of the BC00 themselves can be found in (Bruzual \& Charlot
1993).  Our choice of models assumes that the stars have a Salpeter
(1955) initial-mass function (IMF) slope $\frac{d(log N)}{d(log
M)}=-2.35$.  The lower mass cutoff is $0.1M_{\odot}$ and the upper
mass cutoff is $125M_{\odot}$; these limits (particularly the lower
mass cutoff) affect the associated $M/L_V$ ratios, and thus the
cluster mass estimates.  For each metallicity, the models span ages
from 1 Myr to 15 Gyr.

In order to fit the observed spectral energy distribution of the
clusters with the models, we use a standard $\chi^2$ minimization
technique, where we
fit the age and reddening of the cluster simultaneously.  For each
BC00 model age, we compare the SED to the model reddened by $E(B-V)$
values between 0.0 and 2.0 in steps of 0.02.  For every combination of
age/extinction, we fit the model to the observed cluster
SED, where observations in each filter are weighted by the photometric
uncertainty for that particular measurement.  Each model/reddening
combination results in a $\chi^2$ measurement.  The fit with a minimum
value of $\chi^2$ is adopted as the best fit age/$E_{B-V}$ combination.  The
procedure described above was implemented for all clusters with $UBVI$
imaging.

\subsubsection{Cluster Extinction Distributions}

\begin{figure}
\epsscale{0.5}
\plotone{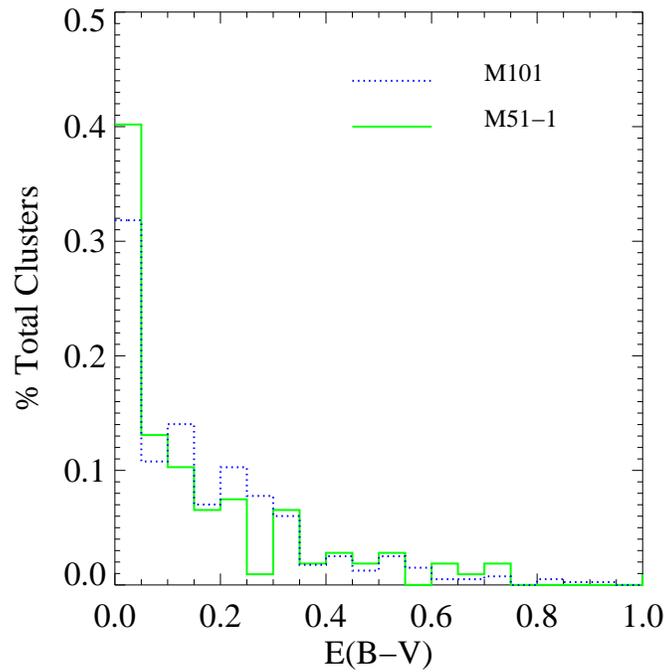}
\caption{The reddening $E_{B-V}$ distribution functions are shown for
our ``all cluster'' catalogs in M101 and M51.  These were
derived using the simultaneous
age/$E_{B-V}$ fitting technique described in $\S3.3.1$.
\label{ebv}}
\end{figure}

In Figure~\ref{ebv}, we show the derived extinction distributions for
all M101-1 and M51-1 clusters, regardless of age.  We will distinguish
between our catalogs containing ``all'' clusters (regardless of age),
and GC candidates, which are a subset of the entire cluster catalog
based on color selections.  The M101 and M51 $E_{B-V}$ distributions
from the age fitting technique described above is (surprisingly)
peaked towards low extinction values ($\sim70$\% have less than 0.1),
confirming the result found by Bik et~al.\  (2003) for M51 clusters.  
This is in sharp contrast to the situation in the Antennae, where we
find that the youngest clusters have a mean $E_{B-V}$ value of 0.9.
Because young clusters dominate our samples and are most likely to
have large reddening, they provide some guidance for typical (upper
limits) for the older clusters in each galaxy.  Thus it appears
unlikely that our ancient cluster samples have significant
contamination from highly reddened young clusters.

\subsection{Final Globular Cluster Selection}

\begin{deluxetable}{rllccr}             
\tablewidth{0pc}             
\tablecaption{GLOBULAR CLUSTER CANDIDATES IN M81}            
\small
\tablehead{             
\colhead{\#} & \colhead {V\tablenotemark{a} }  & 
\colhead{(V$-$I)} & \colhead{(B$-$V)}  & 
\colhead{(U$-$B)}  &
\colhead{$r_\mathrm{eff}$}  
\\
\colhead{} &
\colhead{(mag)} &
\colhead{(mag)} &
\colhead{(mag)} &
\colhead{(mag)} &
\colhead{(pc)} 
}             
\startdata           
1 & $20.649\pm0.052$  & $1.337\pm0.030$  & $1.020\pm0.098$ & $0.117\pm0.395$   & 1.5   \\
2 & $19.824\pm0.011$ & $1.286\pm0.012$  & $ 0.957\pm0.032$ & $0.035\pm0.093$   & 4.0   \\
3 & $21.336\pm0.061$  & $1.250\pm0.050$  & $0.847\pm0.084$ & $0.073\pm0.309$ & 2.7     \\
4 & $21.377\pm0.052$  & $1.530\pm0.026$  & $1.200\pm0.076$ & $0.399\pm0.450$ & 1.5  \\
5 & $20.504\pm0.023$  & $1.125\pm0.019$  & $1.018\pm0.044$ & $0.105\pm0.152$ & 1.0  \\
6 & $19.980\pm0.016$  & $1.197\pm0.023$  & $0.955\pm0.050$ & $0.118\pm0.175$ & 14.2 \\
7 & $22.073\pm0.086$  & $1.226\pm0.060$  & $0.963\pm0.123$ & $0.268\pm0.694$   & 4.4  \\
8 & $21.045\pm0.029$  & $1.175\pm0.026$  & $0.920\pm0.061$ & $0.082\pm0.233$   & 2.6  \\
9 & $20.163\pm0.014$  & $1.376\pm0.014$  & $1.104\pm0.050$ & $0.317\pm0.137$   & 2.1 \\
10 & $19.694\pm0.011$  & $1.057\pm0.017$  & $0.854\pm0.044$ & \llap{$-$}$0.149\pm0.074$   & 0.9  \\
11 & $19.716\pm0.008$ &  $1.185\pm0.015$ & $0.905\pm0.017$ & $0.119\pm0.136$   & 2.6 \\
12 & $20.057\pm0.018$  & $1.089\pm0.014$ & $1.029\pm0.016$ & \nodata & 1.0  \\
13 & $20.432\pm0.023$  & $1.175\pm0.019$  & $0.884\pm0.019$ & $0.097\pm0.124$ & 1.3 \\
14 & $20.371\pm0.018$  & $1.287\pm0.018$  & $1.021\pm0.021$ &  $0.009\pm0.142$ & 1.4 \\
15 &  $19.751\pm0.012$ & $1.325\pm0.018$  & $1.083\pm0.022$ & $0.460\pm0.246$  & 11.5 \\
16 & $19.347\pm0.005$  & $1.116\pm0.008$  & $0.843\pm0.012$ & \llap{$-$}$0.160\pm0.073$ & 1.8 \\
17 & $19.469\pm0.005$  & $1.402\pm0.011$  & $1.003\pm0.018$ & $0.440\pm0.204$ & 9.6 \\
18 & $18.553\pm0.005$  & $1.396\pm0.005$  & $1.139\pm0.008$  & $0.390\pm0.051$   & 2.0 \\
19 & $17.580\pm0.002$  & $1.195\pm0.003$  & $0.766\pm0.004$ & \llap{$-$}$0.127\pm0.020$    & 7.2 \\
20 &  $19.003\pm0.005$ &  $1.142\pm0.006$ & $0.927\pm0.010$ & \llap{$-$}$0.123\pm0.050$    & 3.7 \\
21 & $19.281\pm0.009$  & $1.343\pm0.008$  & $0.992\pm0.010$ & $0.320\pm0.064$   & 1.1 \\
22 & $19.805\pm0.013$  & $1.200\pm0.011$  & $1.002\pm0.015$  & $0.085\pm0.085$   & 2.3 \\
23 & $20.536\pm0.025$  & $1.447\pm0.035$ & $0.965\pm0.040$ & $0.587\pm0.365$   & 3.6 \\
24 & $22.164\pm0.029$ & $1.397\pm0.034$  & $1.017\pm0.060$ & \nodata &  1.1   \\
25 & $20.942\pm0.012$  & $1.305\pm0.020$ & $1.013\pm0.033$ & \nodata &  2.7  \\
26 & $20.145\pm0.008$  & $1.301\pm0.011$  & $1.132\pm0.017$ & \nodata & 2.2 \\
27 & $21.013\pm0.023$ &  $1.188\pm0.024$ &  $0.961\pm0.030$  & \nodata & 3.5 \\
28 & $20.684\pm0.016$ &  $1.414\pm0.016$ &  $1.086\pm0.024$  & \nodata & 1.4 \\
29 & $19.321\pm0.005$ &  $1.151\pm0.008$ &  $0.882\pm0.019$ &  $0.075\pm0.041$ & 1.3 \\
30 & $19.826\pm0.008$ &  $1.521\pm0.010$ &  $1.176\pm0.027$ &  $0.344\pm0.077$ & 0.9 \\
31 & $20.247\pm0.010$ &  $1.089\pm0.016$ &  $0.794\pm0.035$ &  $0.083\pm0.082$ & 6.5 \\
32 & $18.763\pm0.004$ &  $1.234\pm0.006$ &  $0.986\pm0.014$ &  $0.237\pm0.031$ & 0.8 \\
33 & $20.790\pm0.011$ & $1.324\pm0.036$ & $1.245\pm0.252$ & \nodata & 19.8 \\
34 & $19.997\pm0.007$ & $1.138\pm0.009$ & $0.849\pm0.024$ & \nodata & 2.2 \\
35 & $21.903\pm0.031$ & $1.267\pm0.038$ & $1.258\pm0.210$ & \nodata & 9.7 \\
36 & $21.277\pm0.019$ &  $1.307\pm0.044$ &  $0.811\pm0.139$ & \nodata & 8.7   \\
37 & $22.389\pm0.045$ &  $1.193\pm0.053$ &  $0.893\pm0.176$ & $-0.123\pm0.417$   & 1.1 \\
38 & $20.672\pm0.016$ &  $1.302\pm0.020$ &  $1.054\pm0.053$ &  $0.210\pm0.149$   & 7.5 \\
39 & $22.230\pm0.060$ &  $1.294\pm0.052$ &  $1.164\pm0.171$ &  $0.694\pm1.315$   & 3.0 \\
40 & $20.560\pm0.021$ &  $1.795\pm0.021$ &  $1.527\pm0.095$ &  $0.005\pm0.168$   & 3.4 \\
41 & $21.006\pm0.022$ &  $1.179\pm0.033$ & $0.808\pm0.073$ &  $1.056\pm1.306$   & 23.4 \\ 
42 & $20.894\pm0.022$ &  $1.502\pm0.026$ &  $1.130\pm0.076$ &  $0.349\pm0.283$   & 7.7 \\
43 & $22.452\pm0.048$ &  $1.397\pm0.054$ & $0.912\pm0.086$ & \nodata & 4.3  \\ 
44 & $20.360\pm0.010$ &  $1.932\pm0.014$ &  $2.047\pm0.041$ & \nodata & 1.6 \\
45 & $21.941\pm0.016$ &  $1.679\pm0.024$ & \nodata & \nodata & 4.6 \\ 
46 & $22.922\pm0.058$ &  $1.318\pm0.054$ & \nodata & \nodata & 3.1 \\
47 & $22.031\pm0.029$ & $1.306\pm0.035$ & $0.820\pm0.065$ & \nodata & 5.7 \\
 \tablenotetext{a}{V magnitude measured in a $1.0\arcsec$ radius aperture}
\normalsize
\enddata
\end{deluxetable}

For the final globular cluster selection, we reran our age fitting
routine using two additional (lower metallicity) BC00 models: 1/5
solar and 1/50 solar metallicity.  Clusters which were best fit by
ages $\sim9.4$~(log) yrs and older in {\it any} of these models were
selected as globular cluster candidates.  Based on the results of the
SED fitting technique, we find that the following colors can be used
as a reasonable selection criterion for globular clusters when UBVI
photometry is available: $V-I \geq 0.7$, $B-V \geq 0.55$, and $U-B
\geq -0.15$, although there is some variation in the exact values,
depending upon the actual metallicity of the cluster.
If only $BVI$ photometry is
available, we use a slightly more stringent color combination $V-I
\geq 0.8$, and $B-V \geq 0.55$, and if only $VI$ is available (this is
the case for only one field in the halo of M81), we use $V-I \geq
0.8$.  Many of the fields used in this study revealed almost no
background galaxies, suggesting that these spiral disks are relatively
opaque (M81 is an exception).  Because we were able to eliminate the
few observed galaxies on the basis of their morphology, we did not
make a color cut at the red end.

In the next section we quantify the expected contamination from
inclusion of reddened young clusters in fields with only $BVI$
photometry.  Note that for the objects discussed here to actually be
reddened young clusters rather than ancient star clusters, their
$E(B-V)$ values would have to be between $\sim0.4-0.8$,
which is found for extremely few resolved objects in our ``all
cluster'' catalogs.  Finally, the location of the GC candidates were
visually inspected to make sure they did not fall in the center of a
spiral arm, which would significantly increase the probability that a
given object could be a reddened YMC rather than ancient GCs.  Three
such candidates (with BVI photometry) were removed from our ancient
cluster catalog in M51.  Our final GC catalogs, along with photometric
measurements are presented in Tables $3-7$.

\begin{deluxetable}{rllcrc}             
\tablewidth{0pc}             
\tablecaption{GLOBULAR CLUSTER CANDIDATES IN M83}            
\small
\tablehead{             
\colhead{\#} & \colhead{V}  & 
\colhead{(V$-$I)} & 
\colhead{(U$-$V)}  &
\colhead{$r_\mathrm{eff}$} 
\\
\colhead{} &
\colhead{(mag)} &
\colhead{(mag)} &
\colhead{(mag)} &
\colhead{(pc)} 
}             
\startdata           
1 & $20.833\pm0.034$ &  $1.157\pm0.048$ &  $0.738\pm0.121$ & 1.2 \\
2 & $19.666\pm0.016$ &  $1.135\pm0.027$ &  $0.937\pm0.059$ & 2.8 \\
3 & $18.266\pm0.007$ &  $1.140\pm0.010$ &  $1.236\pm0.031$ & 1.0 \\
4 & $17.258\pm0.004$ &  $1.065\pm0.006$ &  $0.943\pm0.016$ & 1.1 \\
5 & $19.217\pm0.010$ &  $1.800\pm0.012$ &  $2.583\pm0.128$ & 1.1 \\
6 & $16.383\pm0.005$ &  $1.441\pm0.006$ &  $1.363\pm0.016$ & 2.1 \\
7 & $21.094\pm0.031$ &  $1.436\pm0.038$ &  $1.633\pm0.328$ & 2.3 \\
8 & $21.798\pm0.050$ &  $1.441\pm0.062$ &  $1.854\pm0.793$ & 6.4 \\
9 & $21.005\pm0.037$ &  $1.785\pm0.044$  & $2.947\pm0.801$ & 2.4 \\
10 & $22.069\pm0.067$ &  $1.167\pm0.093$ &  $1.674\pm0.703$ & 1.9 \\
11 & $21.653\pm0.050$ &  $1.748\pm0.059$ &  $2.063\pm0.662$ & 1.9 \\
12 & $21.042\pm0.032$ &  $1.148\pm0.043$ &  $0.861\pm0.164$ & 4.8 \\
13 & $20.410\pm0.029$ &  $1.019\pm0.043$ &  $1.215\pm0.149$ & 11.4 \\
14 & $20.956\pm0.034$ &  $1.265\pm0.044$ &  $1.312\pm0.221$ & 6.6 \\
15 & $20.717\pm0.034$ &  $1.152\pm0.047$ &  $1.284\pm0.168$ & 4.3 \\
16 & $20.247\pm0.026$ &  $1.419\pm0.033$ &  $2.430\pm0.284$ & 4.3 \\
17 & $20.772\pm0.042$ &  $1.400\pm0.053$ &  $1.076\pm0.149$ & 4.6 \\
18 & $21.033\pm0.034$ &  $1.359\pm0.047$ &  $1.348\pm0.206$ & 1.8 \\
19 & $21.535\pm0.046$ &  $1.062\pm0.060$ &  $1.092\pm0.341$ & 5.0 \\
20 & $19.295\pm0.011$ &  $1.335\pm0.014$ &  $0.978\pm0.051$ & 1.7  \\
21 & $20.704\pm0.025$  & $1.318\pm0.031$ &  $1.316\pm0.161$ & 8.1 \\
\normalsize
\enddata             
\end{deluxetable}

\textit{HST} studies of GC systems in ellipticals, lenticulars, and edge-on
spiral galaxies suggest possible variation in the intrinsic GC color
beyond that seen in the M31 and MW systems.  For example,
Goudfrooij et~al.\  (2003) detected cluster candidates in the halos of
edge-on spirals with significantly bluer colors. They find that 
NGC~4517 has a relatively large number of GC candidates with $0.3 \leq
V-I \leq 0.6$; spectroscopy is needed to confirm whether these are
actually ancient clusters associated with the host galaxy.  Such
objects would not be retained as cluster candidates in our study, as
their colors imply a significantly younger age.

\begin{deluxetable}{rllccc}             
\tablewidth{0pc}             
\tablecaption{GLOBULAR CLUSTER CANDIDATES IN NGC~6946}            
\small
\tablehead{             
\colhead{\#} & \colhead{V}  & 
\colhead{(V$-$I)} & \colhead{(B$-$V)}  & 
\colhead{(U$-$B)}  &
\colhead{$r_\mathrm{eff}$}
\\
\colhead{} &
\colhead{(mag)} &
\colhead{(mag)} &
\colhead{(mag)} &
\colhead{(mag)} &
\colhead{(pc)} 
}             
\startdata           
1 & $20.854\pm0.019$ &  $1.556\pm0.024$ &  $0.951\pm0.045$ &  $0.119\pm0.130$ & 2.3 \\
2 & $22.245\pm0.045$ &  $1.768\pm0.056$ &  $1.063\pm0.140$ &  $0.360\pm0.674$ & 2.6 \\
3 & $21.104\pm0.0204$ &  $1.895\pm0.023$ &  $1.391\pm0.074$ &  $0.199\pm0.243$ & 3.1 \\
4 & $21.771\pm0.029$ &  $1.717\pm0.034$ &  $1.265\pm0.116$  & \nodata & 1.3 \\
5 & $21.367\pm0.023$ &  $1.777\pm0.028$ &  $1.337\pm0.092$  & \nodata & 3.9 \\
6 & $22.000\pm0.020$ &  $1.524\pm0.025$ &  $1.237\pm0.050$    & \nodata & 2.4 \\
7 & $17.688\pm0.003$ &  $1.562\pm0.004$ &  $1.170\pm0.006$   & \nodata & 1.4 \\
8 & $23.136\pm0.044$ &  $1.442\pm0.054$ &  $1.149\pm0.107$    & \nodata & 2.0 \\
9 & $22.162\pm0.022$ &  $1.330\pm0.028$ &  $1.023\pm0.048$   & \nodata & 2.6 \\
10 & $23.452\pm0.053$ &  $1.575\pm0.065$ &  $1.004\pm0.123$   & \nodata & 2.9 \\
11 & $23.150\pm0.038$ &  $1.509\pm0.047$ &  $1.367\pm0.108$   & \nodata & 7.3 \\
12 & $23.183\pm0.040$ &  $1.338\pm0.051$ &  $0.978\pm0.092$    & \nodata & 8.9 \\
13 & $22.196\pm0.028$ &  $1.559\pm0.034$ &  $1.268\pm0.067$   & \nodata & 8.0 \\
14 & $22.770\pm0.057$ &  $1.725\pm0.065$ &  $1.263\pm0.131$   & \nodata & 3.5 \\
15 & $23.141\pm0.069$ &  $1.383\pm0.088$ &  $1.027\pm0.141$  & \nodata & 1.0 \\
16 & $21.687\pm0.027$ &  $1.546\pm0.033$ &  $1.252\pm0.057$   & \nodata & 1.3 \\
17 & $23.394\pm0.082$ &  $1.471\pm0.101$ &  $0.979\pm0.159$   & \nodata & 1.9 \\
18 & $23.009\pm0.041$ &  $1.494\pm0.051$ &  $1.025\pm0.090$   & \nodata & 8.6 \\
19 & $21.151\pm0.013$ &  $1.290\pm0.017$ &  $1.027\pm0.028$   & \nodata & 3.0 \\
\normalsize
\enddata             
\end{deluxetable}

\subsection{Completeness and Contamination Estimates}

\subsubsection{Contamination}

Potential contaminants to our globular cluster catalogs are:
individual stars, background galaxies, blends, and reddened young
massive clusters.  We have eliminated individual stars by requiring
that GC candidates be resolved.  Background galaxies have been mostly
eliminated based on morphology, which is possible with the excellent
resolution provided by HST imaging.  There are two additional reasons
we believe that our cluster samples are essentially free of faint
background galaxies.  The first reason applies to the later-type
spirals in our sample.  In the central portions of these galaxies,
where the density of GCs is expected to be highest, the disks appear
to be nearly opaque.  For example, two of us (BCW and RC) attempted to
locate background galaxies in field M101$-$1, and discovered that
almost no such objects were visible in the entire WFPC2 field of view.
This contrasts with the situation for the earliest-type spiral, M81,
where background galaxies are clearly visible in all fields used in
this work.  However, because few background elliptical galaxies are
expected to be as luminous as the majority of GCs at the distance of
M81, we expect little to no contamination.  This conclusion is
supported by ground-based spectra of M81 GCs selected from these \textit{HST}
fields (from the Chandar, Ford, \& Tsvetanov 2001 catalog), where we
find no background galaxies to $V\sim20$.  Blends and reddened YMCs
may be a more significant problem.  We have eliminated all obvious
blends based on a final visual inspection; however a few closely
blended objects may still remain.

\begin{deluxetable}{rllcrc}             
\tablewidth{0pc}             
\tablecaption{GLOBULAR CLUSTER CANDIDATES IN M101}            
\small
\tablehead{             
\colhead{\#} & 
\colhead {V}  & 
\colhead{(V$-$I)} & 
\colhead{(B$-$V)}  & 
\colhead{(U$-$B)}  &
\colhead{$r_\mathrm{eff}$} 
\\
\colhead{} &
\colhead{(mag)} &
\colhead{(mag)} &
\colhead{(mag)} &
\colhead{(mag)} &
\colhead{(pc)} 
}             
\startdata           
1 & $22.562\pm0.013$ &  $1.070\pm0.060$ &  $0.989\pm0.056$ &  $0.535\pm0.392$ & 3.5 \\
2 & $23.350\pm0.028$ & $1.120\pm0.040$  & $0.919\pm0.114$ &  $0.377\pm0.649$ & 5.0 \\
3 & $23.249\pm0.044$ & $1.148\pm0.056$ &  $1.051\pm0.151$ &  $0.065\pm0.430$ & 3.5 \\
4 & $21.540\pm0.009$ & $1.087\pm0.0130$ & $0.823\pm0.037$ & $0.0350\pm0.086$ & 3.4 \\ 
5 & $23.271\pm0.046$ &  $0.955\pm0.064$ &  $0.518\pm0.102$ &  $0.303\pm0.408$ & 4.4 \\
6 & $23.205\pm0.023$ &  $1.319\pm0.031$ &  $0.927\pm0.107$ &  $0.330\pm1.210$ & 8.3 \\
7 & $23.478\pm0.028$  &  $1.309\pm0.038$ &   $0.726\pm0.113$  & \nodata & 5.0 \\
8 & $20.821\pm0.003$ & $1.895\pm0.004$ & $1.494\pm0.037$ & $0.374\pm0.123$ & 9.4 \\
9 & $22.206\pm0.011$ &  $1.271\pm0.014$ &  $0.889\pm0.055$ &\multicolumn{1}{c}{\nodata} & 3.3  \\
10 & $22.587\pm0.014$ & $1.501\pm0.018$ &  $1.103\pm0.087$ & \multicolumn{1}{c}{\nodata} & 7.5 \\
11 & $22.132\pm0.013$ &  $1.260\pm0.017$ &  $0.686\pm0.050$ &  $0.026\pm0.129$ & 4.8 \\
12 & $23.084\pm0.023$ &  $1.010\pm0.033$ &  $0.602\pm0.083$ &  $0.115\pm0.347$ & 9.2 \\
13 & $22.368\pm0.011$ &  $0.994\pm0.017$ &  $0.549\pm0.046$ &  $0.353\pm0.190$ & 5.9 \\
14 & $20.027\pm0.002$ &  $1.225\pm0.003$ &  $0.859\pm0.016$ &  $0.140\pm0.040$ & 4.1 \\
15 & $20.983\pm0.004$ &  $1.288\pm0.006$ &   $0.894\pm0.027$ &   $0.241\pm0.083$ & 2.5 \\
16 & $23.096\pm0.015$  & $0.774\pm0.030$ &  $0.622\pm0.070$ &  $0.598\pm0.556$ & 3.6 \\
17 & $22.081\pm0.010$ & \phn$1.13\pm0.013$ &  $0.665\pm0.042$ &  $0.228\pm0.142$ & 3.8 \\
18 & $22.456\pm0.008$ &  $0.893\pm0.015$ &  $0.647\pm0.050$ &  $0.156\pm0.164$ & 9.1 \\
19 & $23.642\pm0.027$ &  $0.834\pm0.046$ &  $0.510\pm0.097$ & $-0.042\pm0.287$ & 2.7 \\
20 & $23.893\pm0.028$ &  $1.083\pm0.047$ &  $0.770\pm0.137$  & \multicolumn{1}{c}{\nodata} & 3.7 \\
21 & $21.957\pm0.006$ &  $0.750\pm0.012$ &  $0.524\pm0.036$ &  $0.001\pm0.089$ & 8.0 \\
22 & $23.635\pm0.034$ &  $0.968\pm0.051$ &  $0.517\pm0.111$ & $-0.048\pm0.340$ & 3.5 \\
23 & $23.597\pm0.022$ &  $1.290\pm0.030$ &  $0.723\pm0.118$ &  $0.028\pm0.797$ & 8.7 \\
24 & $23.174\pm0.015$ &  $1.283\pm0.021$ &  $0.621\pm0.080$ &  $0.170\pm0.653$ & 6.3 \\
25 & $23.280\pm0.016$ &  $0.940\pm0.026$ &  $0.623\pm0.086$  & \multicolumn{1}{c}{\nodata} & 3.7 \\
26 & $23.466\pm0.027$ &  $1.185\pm0.036$ &  $0.516\pm0.106$ &  $0.093\pm0.640$ & 6.3 \\
27 & $22.480\pm0.015$ &  $0.978\pm0.021$ &  $0.580\pm0.059$ &  $0.198\pm0.247$ & 6.9 \\
28 & $20.405\pm0.003$ &  $1.599\pm0.004$ &  $1.463\pm0.030$ &  $0.269\pm0.083$ & 6.8 \\
29 & $21.574\pm0.006$ &  $1.184\pm0.008$ &  $0.724\pm0.035$ &   $0.097\pm0.093$ & \llap{1}0.6 \\
\normalsize
\enddata             
\end{deluxetable}

\begin{deluxetable}{rllccr}             
\tablewidth{0pc}             
\tablecaption{GLOBULAR CLUSTER CANDIDATES IN M51}            
\small
\tablehead{             
\colhead{\#} & 
\colhead{V}  & 
\colhead{(V$-$I)} & 
\colhead{(B$-$V)}  & 
\colhead{(U$-$B)}  &
\colhead{$r_\mathrm{eff}$} 
\\
\colhead{} &
\colhead{(mag)} &
\colhead{(mag)} &
\colhead{(mag)} &
\colhead{(mag)} &
\colhead{(pc)} 
}             
\startdata           
1 & $21.271\pm0.022$ &  $0.974\pm0.029$ &  $0.634\pm0.058$ & \llap{$-$}$0.122\pm0.148$ & 12.7 \\
2 & $21.384\pm0.020$ &  $1.045\pm0.028$ &  $0.640\pm0.056$ &  $0.348\pm0.287$ & 5.2 \\
3 & $21.696\pm0.030$ &  $0.889\pm0.040$ &  $0.647\pm0.080$ &  $0.250\pm0.327$ & 6.3 \\
4 &  $21.727\pm0.028$ &  $0.816\pm0.043$ &  $0.623\pm0.080$ &  \llap{$-$}$0.006\pm0.220$ & 9.3 \\
5 & $21.252\pm0.021$ &  $1.092\pm0.027$ &  $0.842\pm0.065$ &  $0.085\pm0.194$ & 8.5 \\
6 & $22.439\pm0.102$ & $ 1.038\pm0.132$ &  $0.707\pm0.196$ &  \nodata & 16.3 \\
7 & $21.357\pm0.041$ &  $1.053\pm0.053$ &  $0.659\pm0.079$ & \llap{$-$}$0.110\pm0.165$ & 4.7 \\
8 & $21.412\pm0.029$ &  $0.973\pm0.038$ &  $0.739\pm0.077$ & \llap{$-$}$0.083\pm0.191$ & 11.3 \\ 
9 & $22.181\pm0.069$  & $0.872\pm0.108$ &   $0.600\pm0.135$ & \nodata & 4.2 \\ 
10 & $22.278\pm0.029$ &  $0.928\pm0.040$ &  $0.584\pm0.052$ & \nodata & 3.7 \\
11 & $20.424\pm0.009$ &   $0.980\pm0.013$ & $0.694\pm0.017$ & \nodata & 3.9 \\
12 & $21.505\pm0.018$ &  $1.339\pm0.022$ &  $1.027\pm0.038$ & \nodata & 5.2 \\
13 & $23.004\pm0.042$ & $1.172\pm0.055$ & $0.756\pm0.082$ & \nodata & 20.7 \\ 
14 & $20.294\pm0.016$ &  $1.147\pm0.020$ &  $0.759\pm0.027$ & \nodata & 5.5 \\
15 & $21.687\pm0.039$ &  $0.776\pm0.059$ &   $0.659\pm0.065$ & \nodata & 7.1 \\
16 & $21.576\pm0.021$ &  $1.090\pm0.028$ &  $0.773\pm0.039$ & \nodata & 13.8 \\
17 & $22.066\pm0.079$ &  $1.078\pm0.109$ &  $0.826\pm0.133$ & \nodata & 7.2 \\
18 & $21.657\pm0.056$ &  $1.114\pm0.067$ &  $0.751\pm0.095$ & \nodata & 9.1 \\
19 & $22.638\pm0.029$ &  $0.878\pm0.042$ &  $0.694\pm0.058$ & \nodata & 12.8 \\
20 & $22.429\pm0.043$ &  $0.841\pm0.058$ &  $0.666\pm0.076$ & \nodata & 9.5 \\
21 & $22.742\pm0.039$ &  $1.067\pm0.053$  & $0.658\pm0.072$ & \nodata & 6.5 \\
22 & $23.097\pm0.036$ &  $0.845\pm0.053$  & $0.559\pm0.070$ & \nodata & 7.1 \\
23 & $21.954\pm0.039$ &  $0.812\pm0.057$ &  $0.677\pm0.063$ & \nodata & 4.4 \\
24 & $21.431\pm0.019$ &  $0.877\pm0.028$ &  $0.576\pm0.033$ & \nodata & 4.3 \\
25 & $22.482\pm0.052$ &  $0.813\pm0.076$ &  $0.664\pm0.085$ & \nodata & 5.7 \\
26 & $22.368\pm0.028$ &  $1.504\pm0.035$ &  $1.197\pm0.066$ & \nodata & 8.7 \\
27 & $21.438\pm0.014$ &  $0.916\pm0.021$ &  $0.709\pm0.028$ & \nodata & 8.4 \\
28 & $22.806\pm0.037$ &  $1.479\pm0.045$ &  $0.975\pm0.078$ & \nodata & 9.5 \\
29 & $21.937\pm0.033$ &  $0.975\pm0.043$ &  $0.590\pm0.069$ & \nodata & 12.0 \\
30 & $22.610\pm0.030$ &  $1.095\pm0.041$ &  $0.738\pm0.058$ & \nodata &  7.5 \\
31 & $23.373\pm0.045$ &  $1.119\pm0.060$ &  $0.583\pm0.090$ & \nodata & 7.0 \\
32 & $23.409\pm0.063$ &  $0.744\pm0.090$ &  $0.632\pm0.125$ & \nodata & 3.6 \\
33 & $23.629\pm0.076$ &  $1.219\pm0.110$ &  $0.913\pm0.141$ & \nodata & 11.9 \\
34 & $20.837\pm0.070$ &  $1.252\pm0.098$ &  $0.839\pm0.100$ & \nodata & 2.3 \\
\normalsize
\enddata             
\end{deluxetable}

Because some of our ancient clusters were selected from $BVI$
photometry (when no U band imaging was available), there is likely
some contamination by reddened young clusters which cannot be
sorted out from ancient objects when only these three filters are
available.  Here, we use the available UBVI imaging in each galaxy to
estimate the number of potential (reddened) young clusters in our
sample.  
We used the following technique:
clusters which would be selected as GCs according to the BVI color
criteria given in $\S3.4$, were compared with the fraction selected
using our UBVI criteria.  The fraction of clusters which are clearly
young and reddened based on UBVI is assumed to hold for the rest of
our GC catalog.  For M51-1, we find that (for objects brighter than
21.7), only 1 out of 7 has colors consistent with a highly reddened
YMC rather than an ancient GC. Since field 1 covers inner and spiral
arm regions, the cluster reddening distribution might reasonably be
expected to be representative for the rest of the galaxy, if not an
overestimate.  Out of 34 total GC candidates in M51, seven have U band
photometry.  If 1/7 of those with only BVI photometry are expected to
be reddened young clusters, we expect $\sim4$ contaminants in our M51
cluster sample.  In M81, an examination of our entire cluster catalog
shows a contamination fraction of $\sim20$\% for our GC sample (mostly
at the faint end), resulting in an estimated $\sim4$ reddened young
cluster contaminants.
In M101, only four of the 29 GC candidates have no U band measurement
(due to faintness).  For the brighter portion of the sample,
we found that $\sim1$ out of 5 clusters which had BVI colors typical
of ancient clusters were actually reddened young clusters.  Thus
statistically 
we expect a maximum of one M101 sample clusters to be young.
Photometry of all clusters in NGC~6946-2 suggests that no young clusters 
are in our GC catalog.

\subsubsection{Completeness}

The completeness of our sample will depend upon a number of complex
issues.  Completeness in terms of cluster size is one issue, since we
have only included resolved objects in this study.  In general, we can
be reasonably confident that an object is extended if its FWHM is
about 0.2 pixels larger than the stellar PSF.  At the target galaxy
distances, this 0.2 pixel lower size limit corresponds to an effective
radius of 0.5, 0.6, 0.9, 1.1, and 1.2~pc for M81, M83, NGC~6946, M101,
and M51 respectively.  This can be compared with the Galactic GC
system to get a very approximate idea of completeness based on size,
{\it if} the GC systems in these galaxies have similar
$r_{eff}-\mbox{galactocentric}$ distance distributions as their Milky
Way counterparts.  We use the McMaster list (Harris 1996) to estimate
the number of Milky Way clusters which would fall out of our samples
based on their compactness and photometric properties.  Seven Galactic
GCs have half mass radii ($r_{\frac{1}{2}}$) smaller than 1.1~pc, and
nine have $r_{\frac{1}{2}} \leq 1.2$.  However, these Galactic GCs
have integrated luminosities of $M_V\sim-4~\mbox{to}-7$, and so are
fainter than the expected GC turnover.  Assuming that any missing,
compact clusters in our target galaxies follow a similar pattern, the
technique used to estimate the total number of GCs (described in
$\S4.5.1$) should not be affected.  Thus, we do not make any
correction for our inability to detect the most compact clusters.

\begin{figure}
\epsscale{0.5}
\plotone{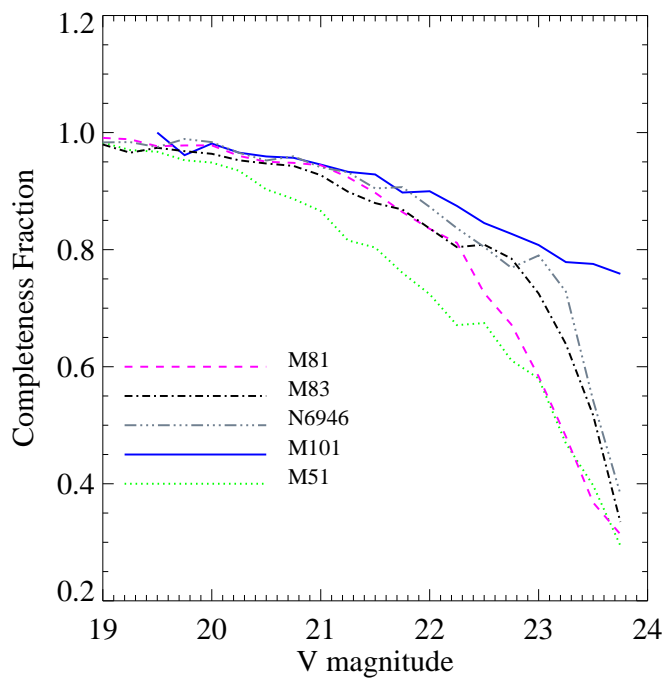}
\caption{Average V band completeness curves as determined from artificial
cluster experiments are shown for each target galaxy.
\label{complete}}
\end{figure}

Because of the complicated and often messy spiral regions, and because
our detection algorithm requires a final ``by eye'' check, it is not
easy to exactly quantify our completeness levels.  We assume that our
detection algorithm recovers all resolved clusters (a thorough and
independent inspection of clusters in M101 by both R.C. and
B.W. suggests that this is a reasonable assumption), even though it
may leave in a few blends.  Artificial cluster experiments were
performed by adding artificial GCs (generated from the ADDSTAR task in
IRAF, where instead of stars, clusters were selected) to our images,
and then rerunning these through the automated portion of our
detection algorithm.  These 'fake' clusters were added in groups of 50
in randomly placed positions on each chip, and then detected and
re-photometered.  We assume that as long as a GC made it through the
automated pipeline, it was not thrown out during the visual inspection
phase (which was used primarily to weed out blends).  In
Figure~\ref{complete} we show average V band completeness functions
for the photometry of GC candidates in each galaxy.  Formal
completeness levels are likely somewhat optimistic, since the
synthetic clusters have been created from previously identified
clusters in each field.  As expected from the total V band exposure
times, the M101 data has a higher completeness level at a given
magnitude than the other target galaxies.

\section{Results: Globular Cluster System Properties}

\subsection{Color and Luminosity Distributions}

\begin{figure}
\epsscale{0.4}
\plottwo{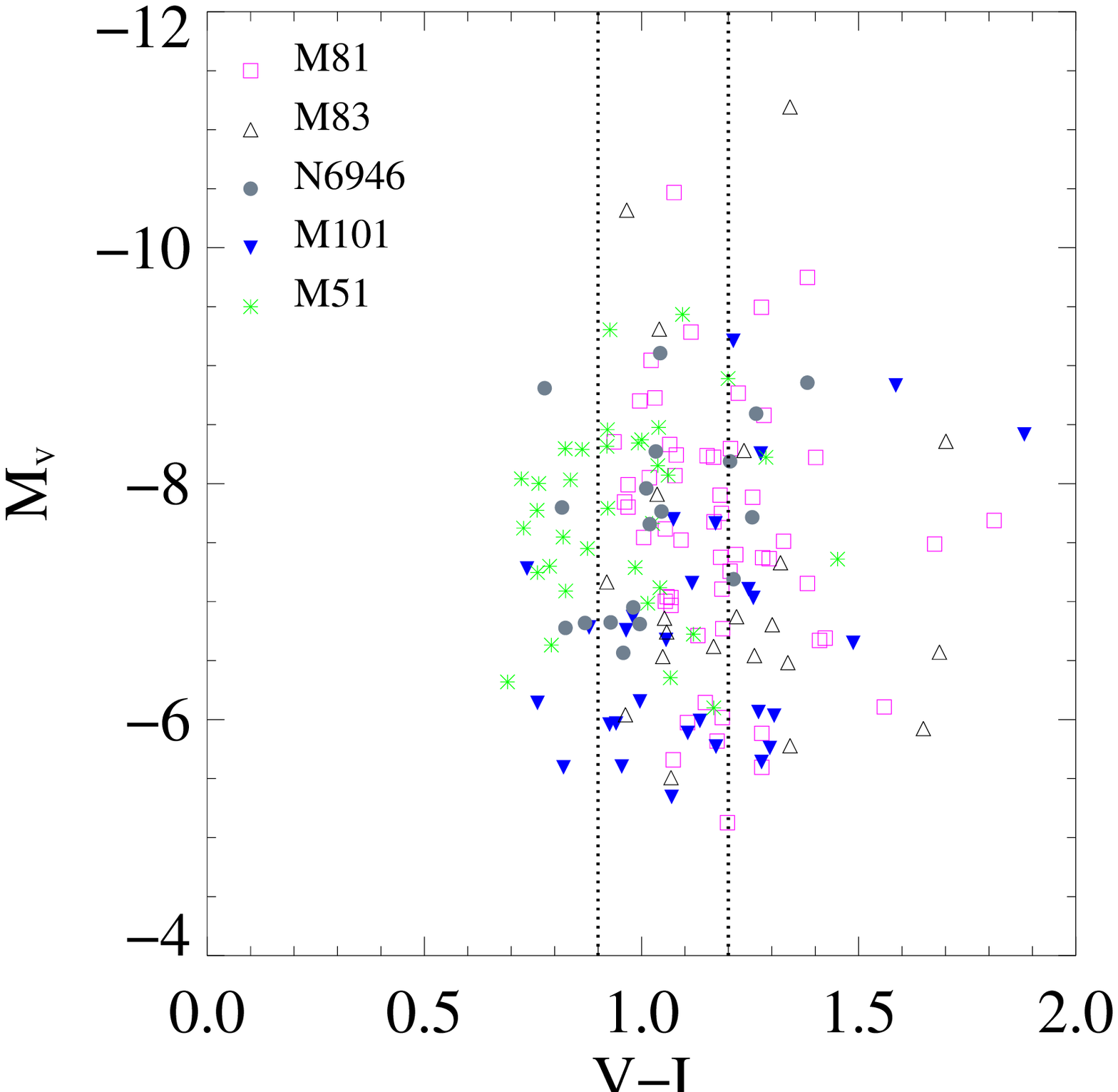}{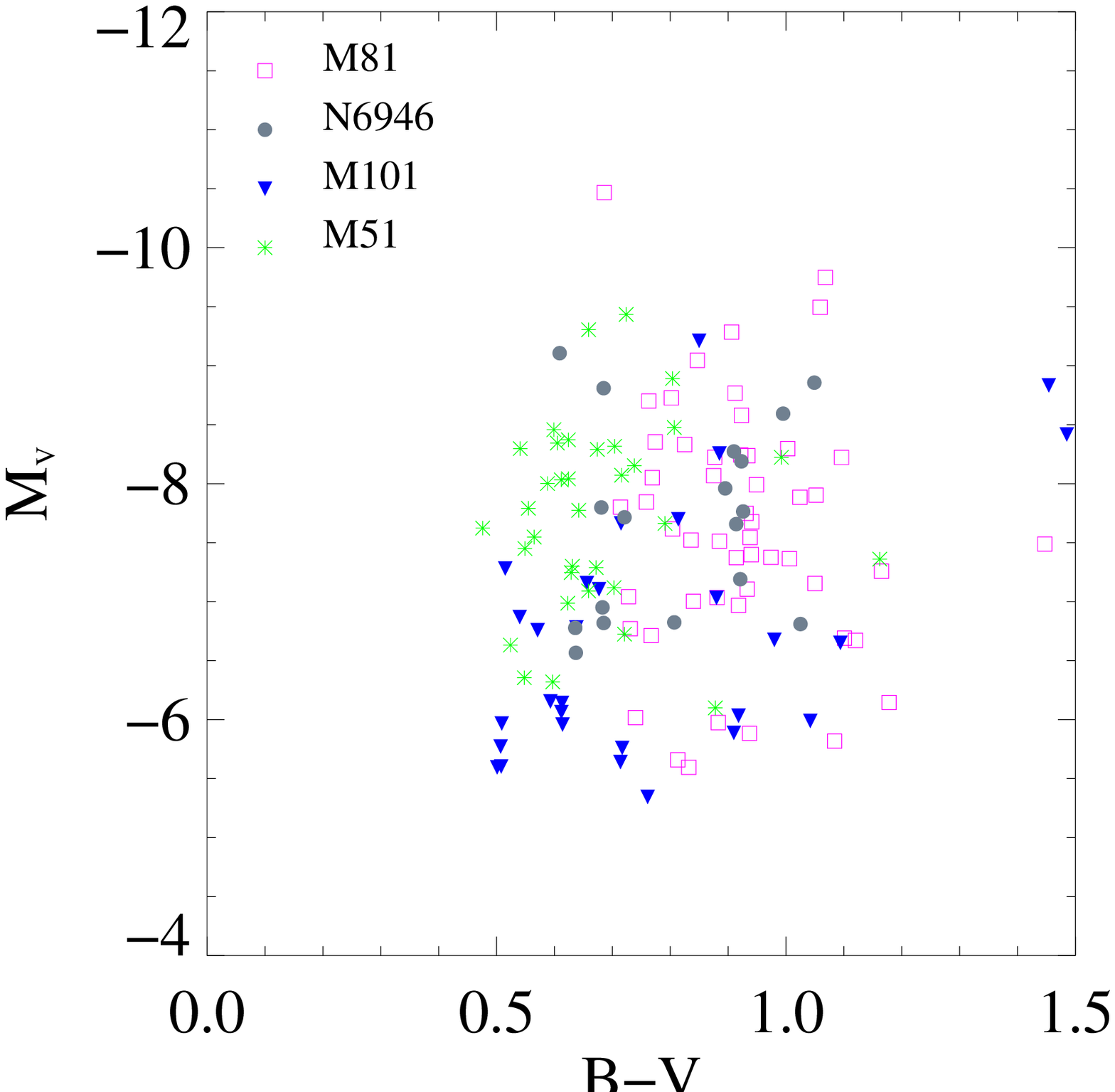}
\caption{The color-luminosity distributions of GCs presented in this
work are shown.  The luminosity has been corrected for distance and
foreground extinction, and the colors have been dereddened by the
foreground value only.  The dotted lines show typical $(V-I)_0$ colors
for blue and red GC subpopulations in ellipticals and lenticulars.
\label{cmd}}
\end{figure}

Figure~\ref{cmd} shows the $(V-I)$ vs. $V$ and $(B-V)$ vs. $V$ color
magnitude diagrams (CMD) of our detected globular cluster samples.
Colors have been dereddened by the foreground extinction values and
magnitudes corrected for foreground extinction and distance.  In
Figure~\ref{cmd}a, we show the mean $(V-I)$ colors of the two peaks
found for GC systems in many elliptical and lenticular galaxies,
at typical values of 0.9 (blue, metal poor) and 1.2 (red, metal rich)
(Kundu \& Whitmore 2001).  Below, we discuss the global luminosity and
color distributions for the GC systems in our spiral sample.

One of the most striking features in Figures~\ref{cmd}a,b is that the
globular clusters from different galaxies appear to separate in color
space.  The M51 cluster population has mean (foreground reddening
corrected) $(B-V)$ and $(V-I)$ colors of 0.67 and 0.95, with standard
deviations of 0.14 and 0.17 respectively.  Comparable values for the
M81 GC sample are 0.92 and 1.19, with standard deviations of 0.15 and
0.18.  The mean $(V-I)$ color of the M51 GC system is remarkably
similar to the blue (metal-poor) peak found in elliptical and
lenticular galaxies (e.g., Burgarella, Kissler-Patig, \& Buat 2001),
and the mean $(V-I)$ color of our M81 GC sample is remarkably similar
to the red (metal-rich) peak in these early type galaxies.  The M83
system has a mean $(V-I)$ color of 1.22 with a standard deviation of
0.23 -- similar to the metal-rich peak in early type galaxies, but
with a large spread.  Note that this is primarily due to a number of
clusters fainter than $M_V\sim-7$, which have predominantly red
colors.

Because colors are sensitive to metallicity in single stellar
populations older than a few Gyr, the GC color distributions seem to
suggest that M51 has a nearly exclusive metal-poor GC population,
despite being of a similar Hubble type as the Milky Way, which is
known to have formed $\sim40-50$ GCs more metal rich than
$\mbox{[Fe/H]}\sim-1$ (see compilation in Harris 1996).  The color
distributions for M81 GCs however, are redder, suggesting the
presence of metal-rich GCs
(although internal reddening would shift any affected cluster to bluer
colors).  The color distributions are explored further in the next
two sections.

\begin{figure}
\centerline{\includegraphics[height=2.5in]{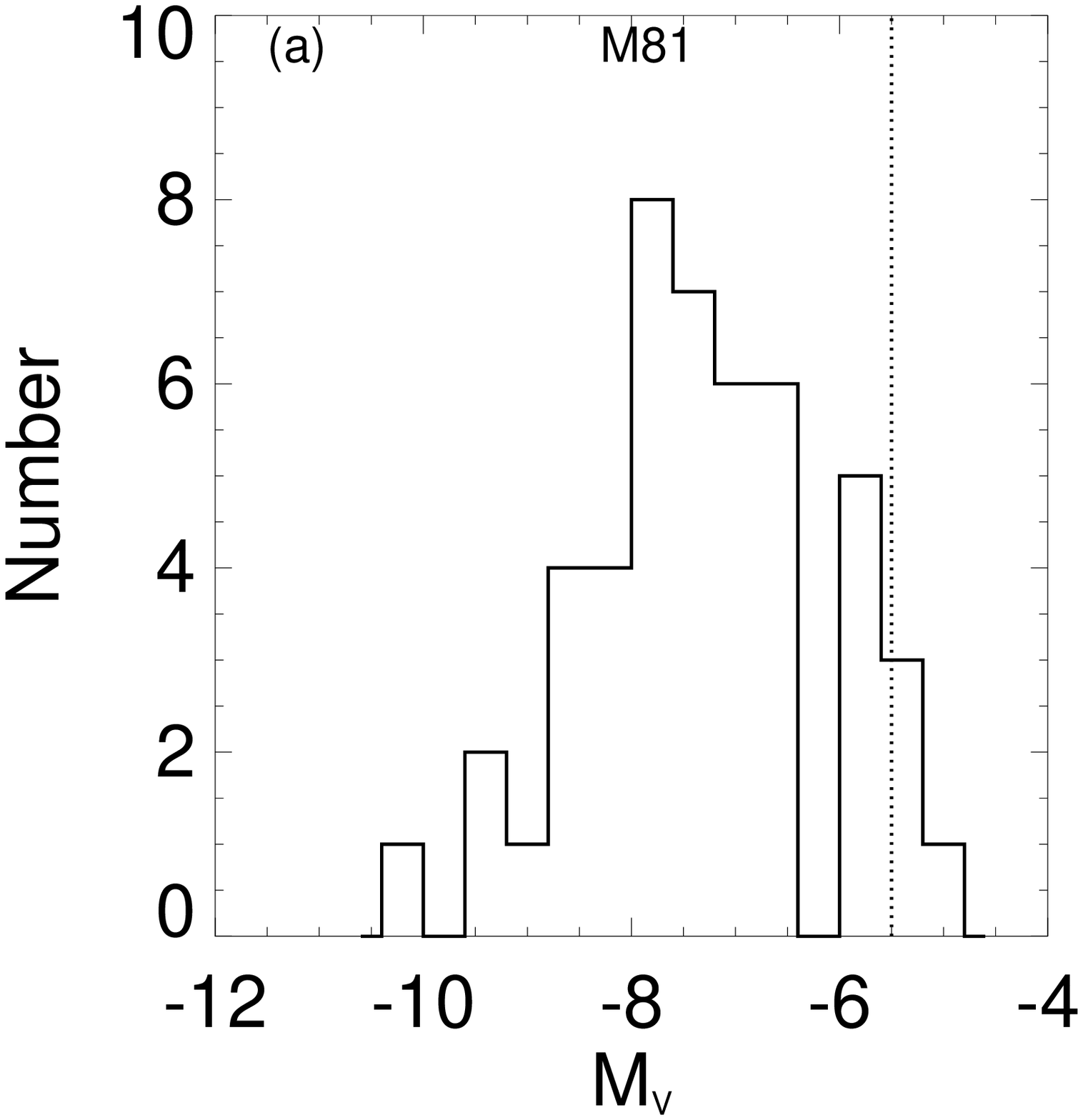}~~\includegraphics[height=2.5in]{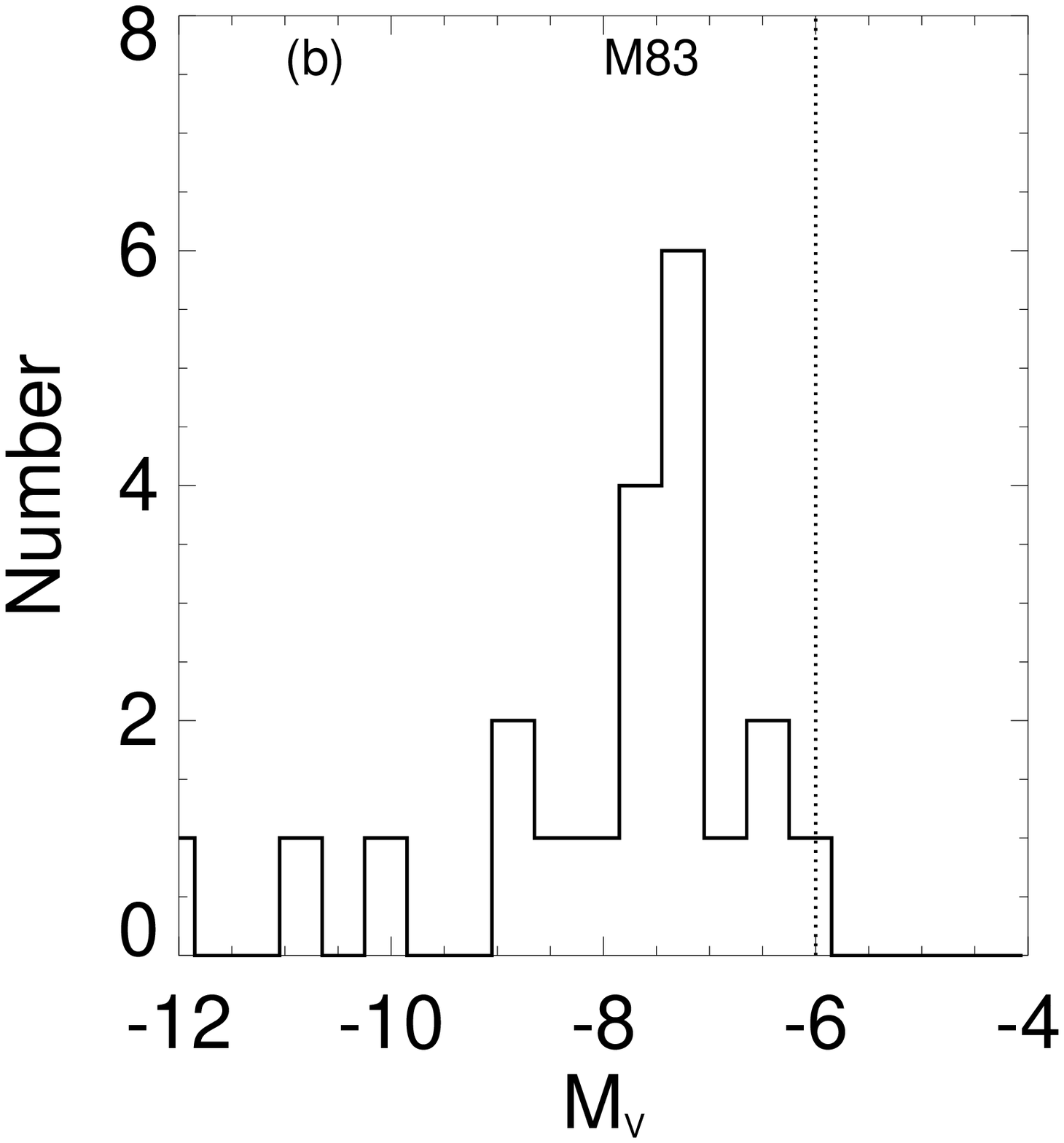}}
\vspace{4pt}
\centerline{\includegraphics[height=2.5in]{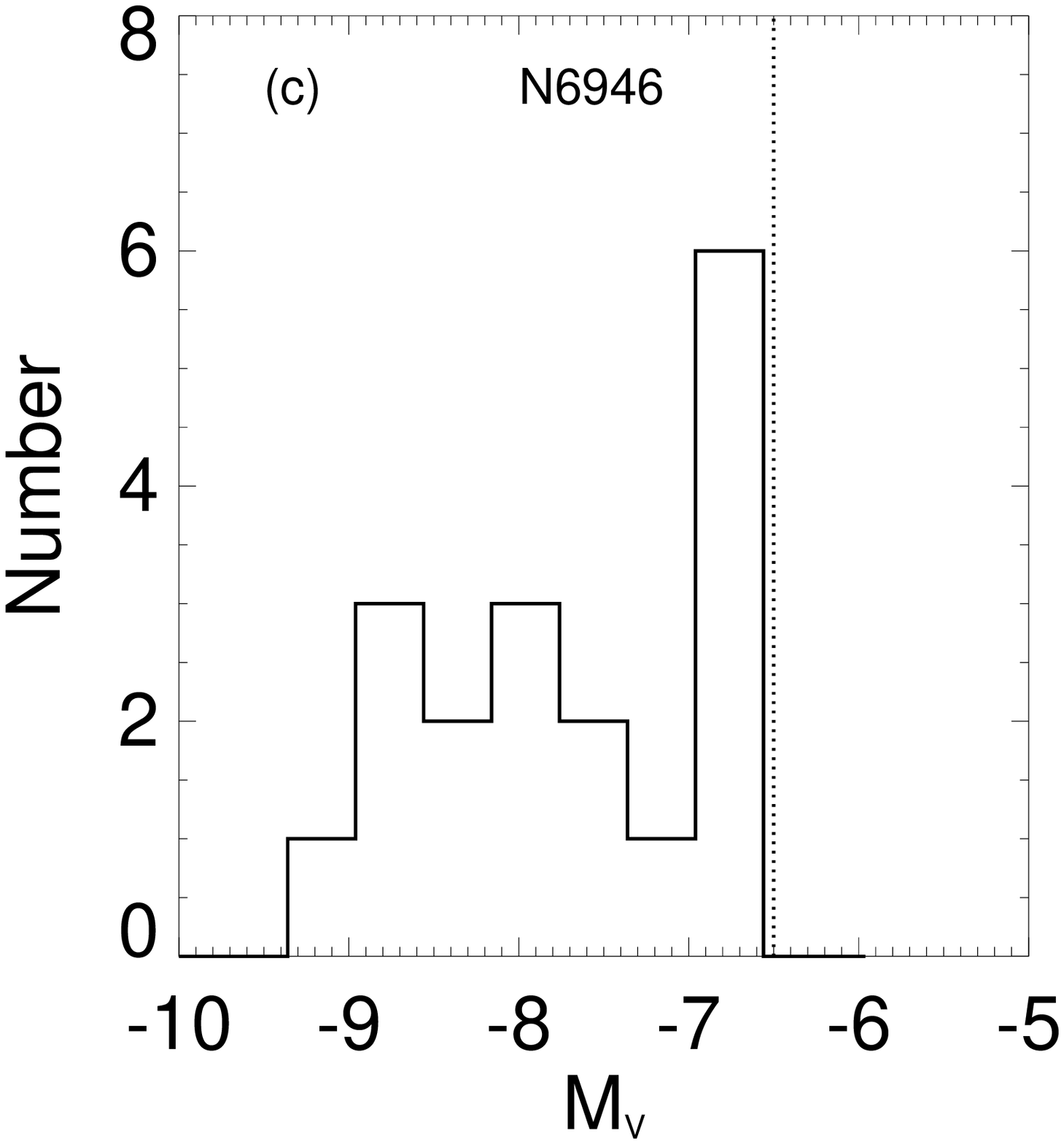}~~\includegraphics[height=2.5in]{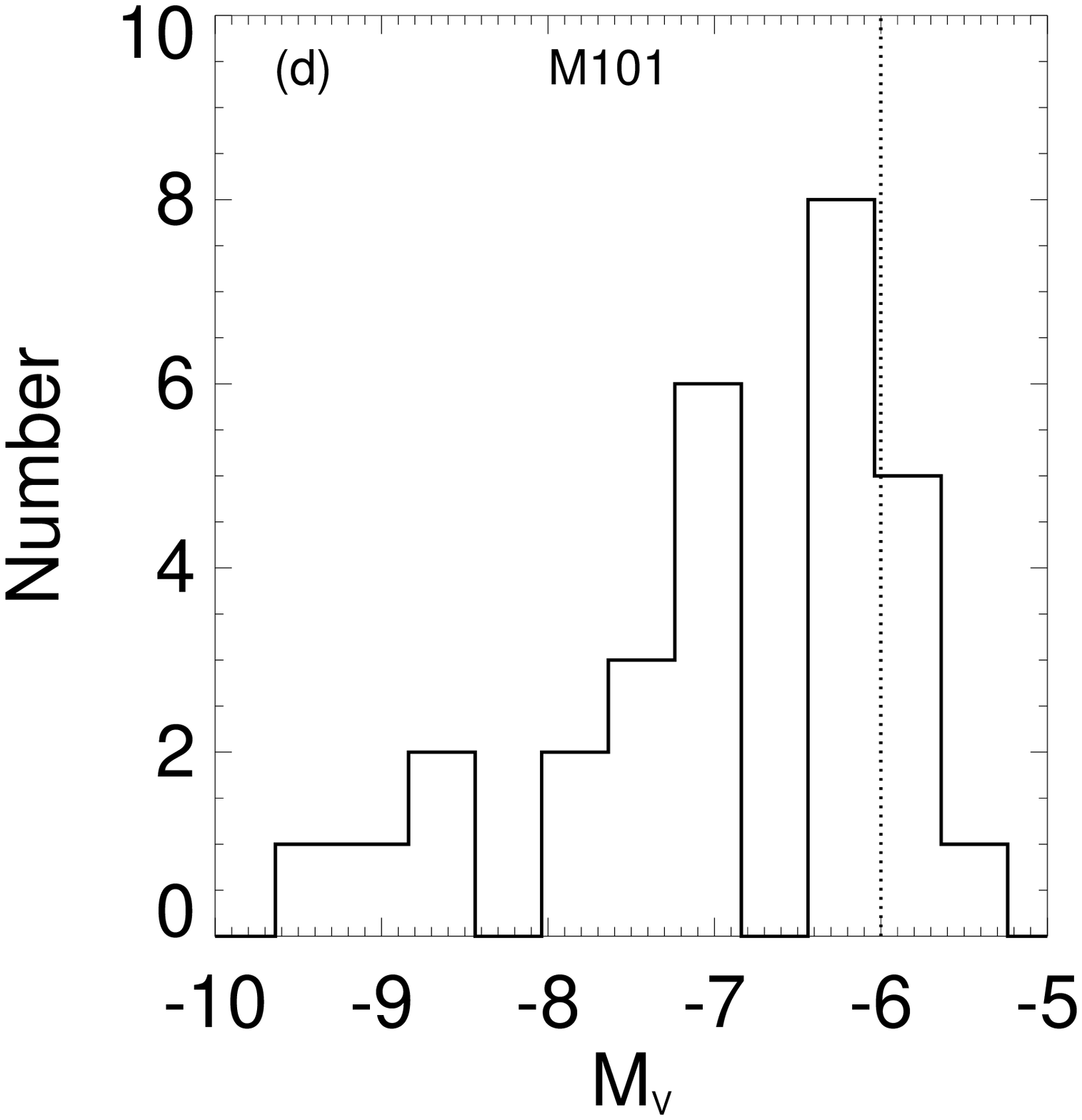}}
\vspace{4pt}
\centerline{\includegraphics[height=2.5in]{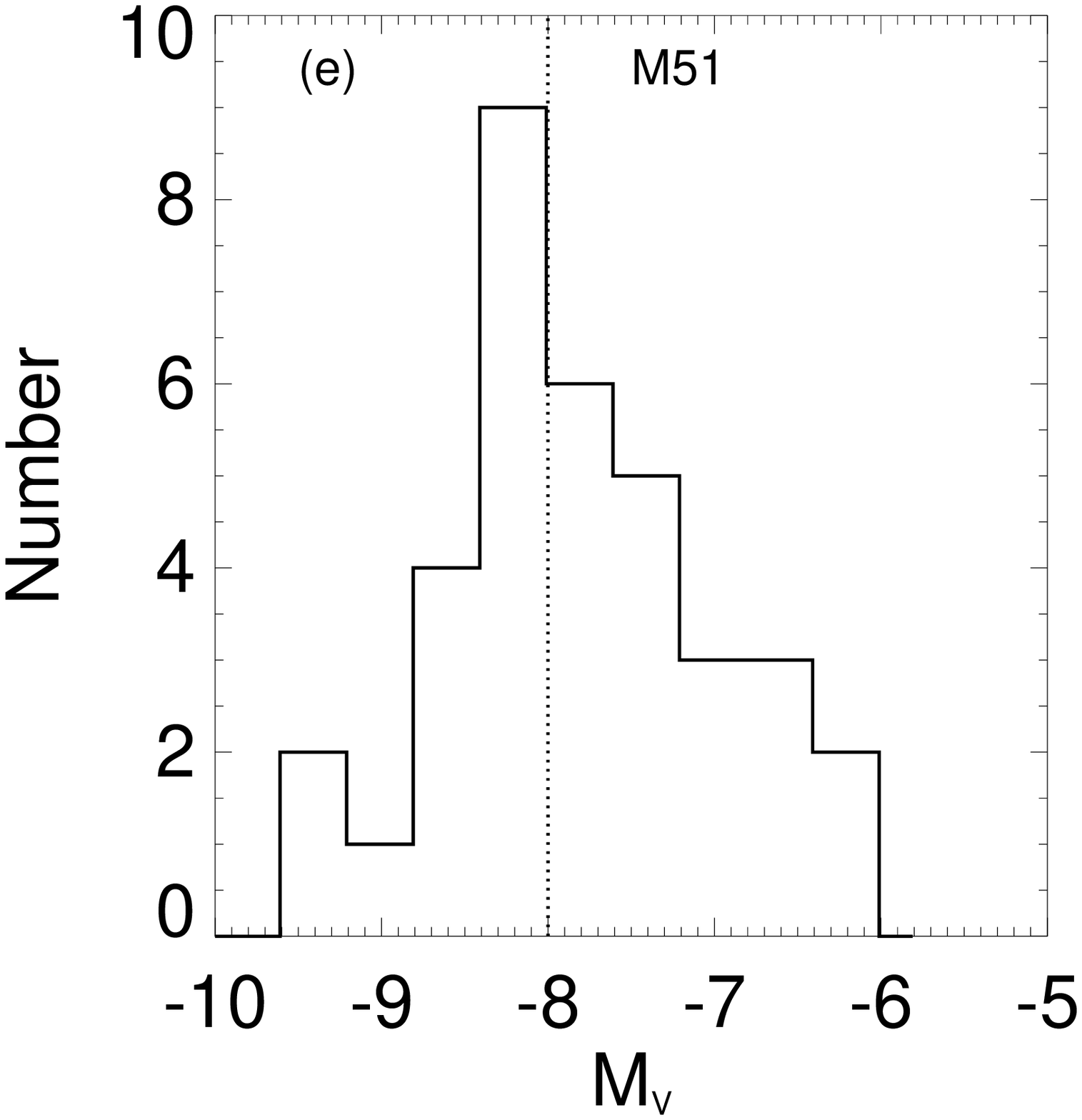}~~\includegraphics[height=2.5in]{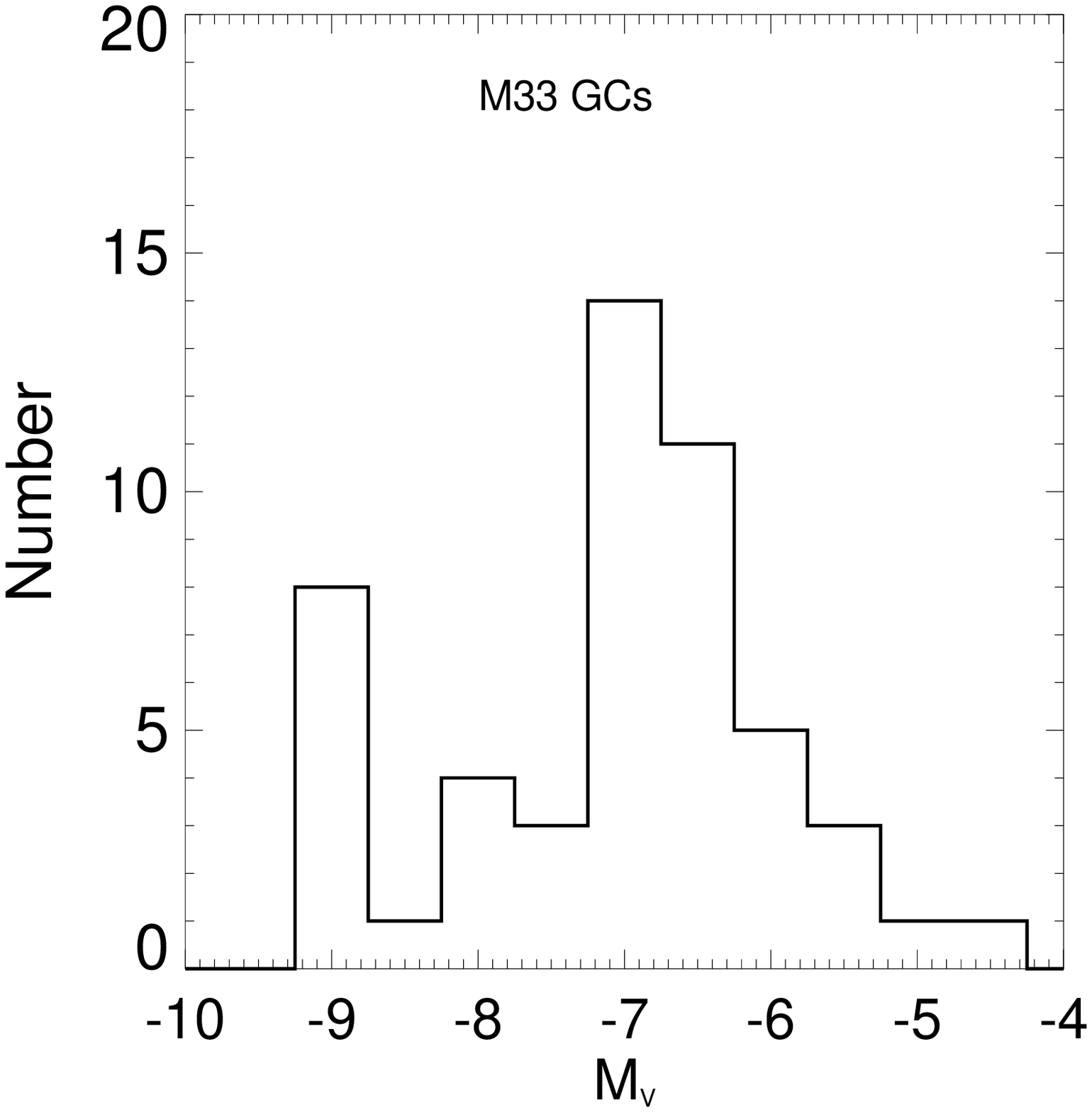}}
\caption{Globular cluster luminosity functions for each galaxy are
plotted.  These have not been corrected for incompleteness.  The
dotted line represents that average 80\% completeness level for each
cluster sample.  For comparison, we have added the luminosity function
for M33 GCs (Chandar et~al.\  2001) in the last panel.
\label{gclf}}
\end{figure}

In Figure~\ref{gclf} we show the observed V band luminosity
distributions for our globular cluster samples, uncorrected for
completeness.  The dashed lines represent average 80\% completeness
limits, as discussed in section 3.5.2.  Note that these are not the
completeness as a function of local background level, and that a
single value for each galaxy cannot capture the complicated issue of
completeness.  For M81, the closest and earliest type spiral in our
sample, we see a peak in the GC luminosity function more luminous than
the completeness level.  This is the characteristic shape and turnover
seen in the Milky Way, M31 and most elliptical and lenticular GC
systems.  Thus M81 GCs appear to have a shape similar to the now
familiar ``universal'' GC luminosity function.  While our M83 sample
does not contain a large number of GCs, the luminosity function for
these objects is similar to that for M81.

M51 is the most distant galaxy in our sample, and does not have
extremely deep exposures.  Therefore the completeness limit for this
galaxy does not quite reach the turnover in the GC luminosity function
(which is expected to occur near $M_V \sim -7.4$).  The apparent peak
in the cluster luminosity distribution around $M_V\sim-8.5$, is likely
caused by one or two effects: 1) variable completeness limits as a
function of background level, or 2) possible contamination from reddened
young clusters, and is likely not real.  

The situation in M101 appears to be quite different from that in
M81. While the number of clusters is few, and based primarily on a
single \textit{HST} WFPC2 pointing located near the center, our deep drizzled
observations reveal a population of faint, red clusters, which appear
to have a powerlaw luminosity distribution down to our completeness
limit ($M_V\sim-6$).  We remind the reader that all of these objects
are {\it resolved}, so cannot be individual stars.  The colors for
these faint objects are indistinguishable from the more luminous
clusters in our sample (although due to their faintness, the U band
photometry has higher uncertainty).  The nature of these faint, red
clusters is discussed further in $\S5.1$.  Although the GC catalog for
NGC~6946 contains relatively few objects, we note that the luminosity
distribution appears more similar to that for M101 GCs rather than M81
GCs, due to the apparent ``excess'' of clusters beyond the expected
turnover of $M_V\sim-7.4$.

Low number statistics may play a role in the observed luminosity
distributions for M101 and NGC~6946.  We quantified this effect by
performing a simple experiment.  A parent gaussian distribution with a
peak at $M_V=-7.4$ and a width, $\sigma=1.2$ (mimicking fits to the
Galactic GC system distribution) was assumed.  We imposed a cutoff of
$M_V=-6$, roughly the 50\% completeness limit for NGC~6946, according
to Figure~3.  This truncated Gaussian was then randomly sampled 19
times, and the resulting distribution of luminosities displayed in a
histogram, similar to those shown in Figure~5.  We find that roughly
$\sim1/3$ of the time, a distribution somewhat similar to the GCLF for
NGC~6946 results, and $\sim2/3$ of the time the distribution has more
clusters near the peak magnitude.  Repeating this experiment using a
collection of 29 clusters and comparing with the distribution plotted
for M101, a similar distribution with excess faint clusters resulted
only $5-10$\% of the time.  Therefore, there is a $90-95$\% probability
that the M101 GC luminosity function differs substantially from
that observed in the Milky Way and a number of other galaxies. 

\subsection{Color-Color Distributions}

The metallicity distributions of GC systems shed light on the
formation history of the parent galaxy.  For the GC systems in
elliptical galaxies, widespread bimodality in the color (and by
extension metallicity) distributions, has helped constrain the most
likely formation scenarios for early-type galaxies (e.g., Kundu \&
Whitmore 2001).  However, less is known concerning the metallicity
distributions of GC systems in spirals.  The two best studied spirals,
the Galaxy and M31, both have bimodal GC metallicity distributions
(e.g., Cote 1999; Perrett et~al.\  2002).

\begin{figure}
\epsscale{0.4}
\plottwo{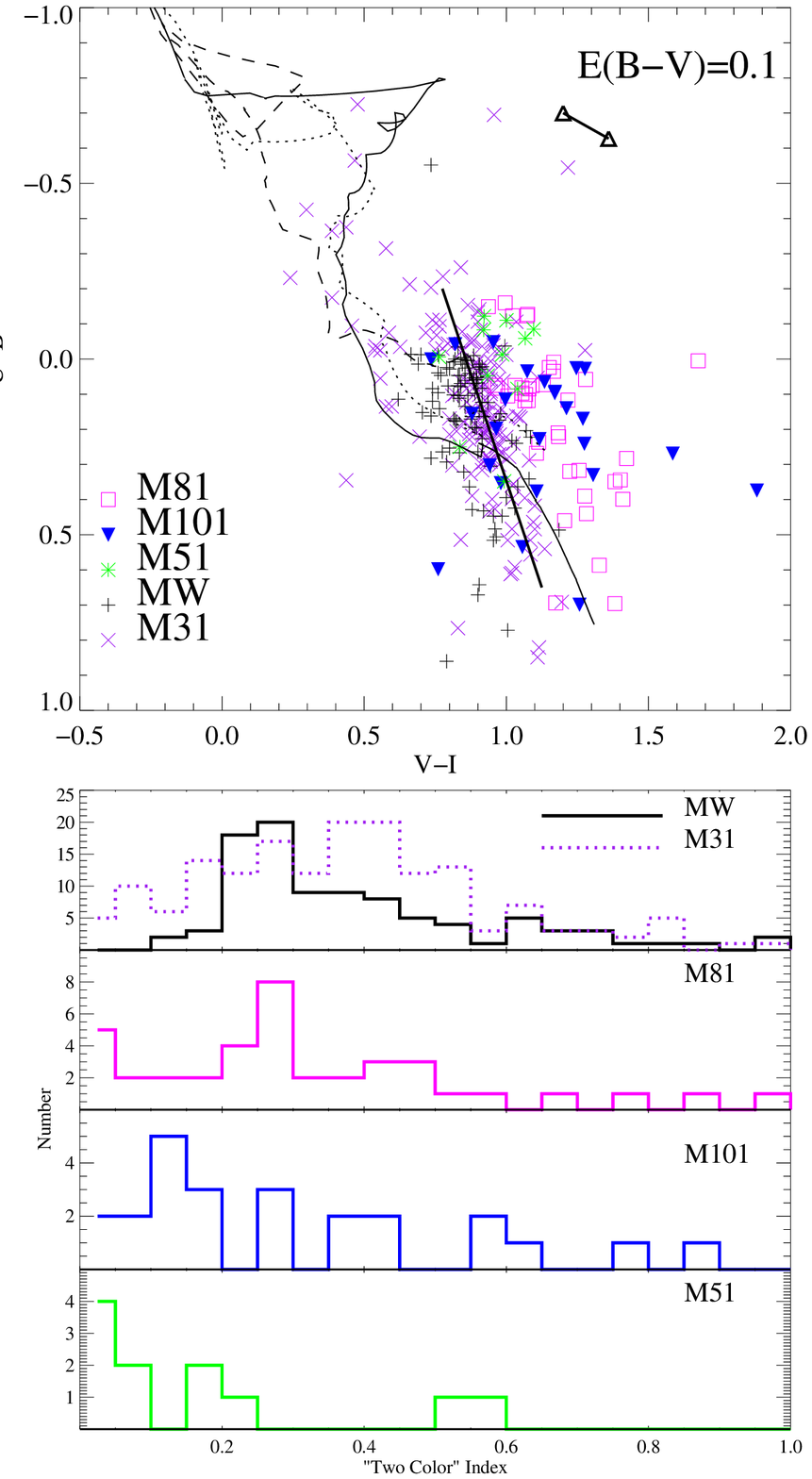}{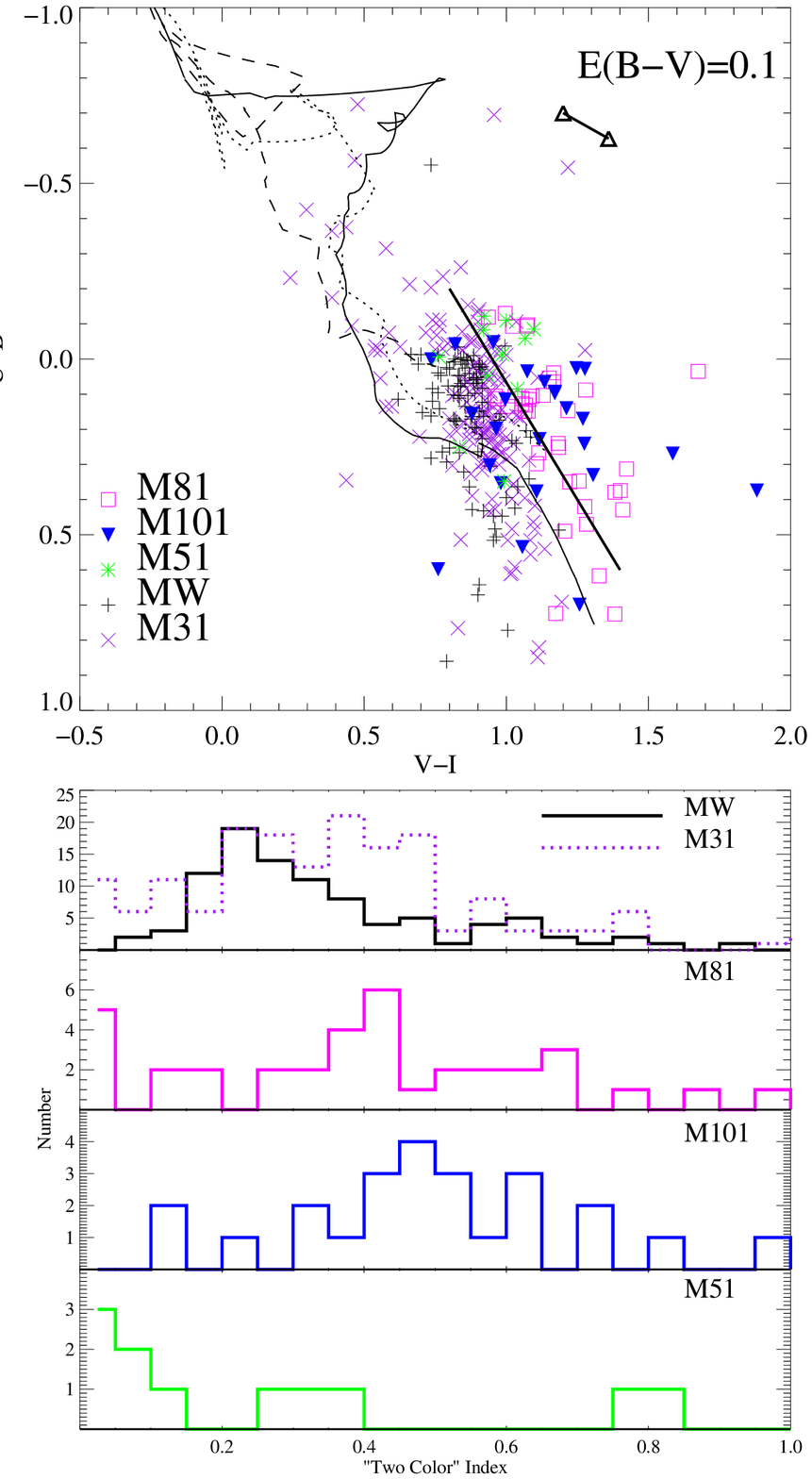}
\caption{{\it Top Panels:} The $(V-I)$ vs. $(U-B)$ color-color diagram
is shown.
The lines are solar (solid), 1/5~solar (dotted), and 
1/50~solar (dashed) BC00 metallicity models.  The direction of the
reddening vector is shown by the arrows.  Milky Way and M31 GC colors
have been dereddened (both foreground and internal).  The GC colors
from this work have been dereddened by only the foreground values.
The solid lines show the best fit to dereddened M31 GCs from the
Barmby et~al.\  (2000) catalog (left), and the best fit to the M81 plus
M51 data points (right).
{\it Bottom Panels:}  The distribution of two color values
(see text for description) are plotted.
These show the extended nature of the M31 and Milky Way GC systems,
as well as those in M81 and M101.
\label{2colmetal1}}
\end{figure}

In Figure~\ref{2colmetal1}, we plot the cluster $(V-I)$ vs. $(U-B)$
color-color distributions.  The GC candidates have been dereddened by
foreground $E_{B-V}$ only.  These are compared with three
different metallicity stellar evolutionary BC00 models: solar (solid
line), 1/5 solar (dotted line) and 1/50 solar (dashed line); clearly
more metal-rich models have redder colors for ancient populations.
For comparison, we also plot the colors of Galactic GCs (Harris 1996)
and $\sim170$ M31 GCs (with UBVI photometry) presented in Barmby et
al. (2000).  These have been corrected for both foreground and
internal extinction (P. Barmby kindly made the derived $E_{B-V}$
values for M31 GCs available to us), for clusters where the $E_{B-V}$
derivation is robust, and by only the foreground value when it is not.
We note that there is some scatter in the M31 GC colors, most notably
from a handful of blue objects.  These are likely young, massive
clusters found in the disk of the Andromeda galaxy, as
spectroscopically confirmed by Barmby et~al.\  (2000).

Blue Galactic GC colors agree well with the models, while the redder
ones appear to have colors which are offset (blueward) from the high
metallicity model predictions of BC00.  The dereddened M31 GC colors
agree well with their Milky Way counterparts.  The M81 GCs presented
here however, appear to lie along a different locus than both Milky
Way and M31 GCs.  Potentially, this is due to internal reddening,
which we have not corrected for.  We find that if M31 GC colors are
only corrected for foreground reddening, they lie in the same region
as the M81 GCs, indicating that internal reddening is a plausible
explanation for the offset.  A second factor supporting the
possibility that differential reddening is responsible for the offset
between M81 and dereddened M31 GC colors is the location of our M81
fields, which are scattered mostly along the spiral arms and disk.
M51 and M101 clusters follow the intrinsic Galactic and M31 GC
color-color locus more closely.

We attempt to use the color-color distributions to study the
underlying metallicity distribution in spiral GC systems in two ways.
First, because the intrinsic colors of Galactic and M31 GCs are in
good agreement, we assume that these provide a fiducial for the
clusters studied in this work.  Figure~\ref{2colmetal1}a shows a
linear fit to intrinsic M31 GC colors in $(V-I)$ vs. $(U-B)$
color-color space [equation: $(U-B)$ = $3.33\times (V-I)$-2.9].  We
assume that deviations from this line are due to internal reddening
for the GCs presented in this work, and track them along the
reddening vector until they intersect the best fit line, assuming the
$R_V=3.1$ extinction curve of Cardelli, Clayton, \& Mathis (1989).  We
note that Barmby et~al.\  (2000) found little difference in the
extinction law between Galactic GCs and their M31 counterparts, and we
assume the Galactic extinction law is also similar for the galaxies
studied in this work.  Once
we dereddened M81, M101, and M51 GCs, we determined the position
of each point along the best fit line;
hereafter we refer to this value as the ``two color index''.  The
dereddened measurements in color-color space of {\it ancient} star
clusters should reflect the underlying metallicity distribution.
Histograms of the two color indices for each galaxy sample are shown
below the color-color diagram.

A second possibility is that the M81 GCs in our spiral sample have
different intrinsic colors than those in the Galaxy and M31.  To
explore this possibility, we fit the locus of the M81 GCs in
Figure~\ref{2colmetal1}b.  We then determined the two color index for
each object by finding the location of the perpendicular bisector for
each cluster.  The resulting histograms for Galactic, M31, M81, M101,
and M51 GCs are shown in the lower panel of Figure~\ref{2colmetal1}b.

\begin{figure}
\epsscale{0.4}
\plottwo{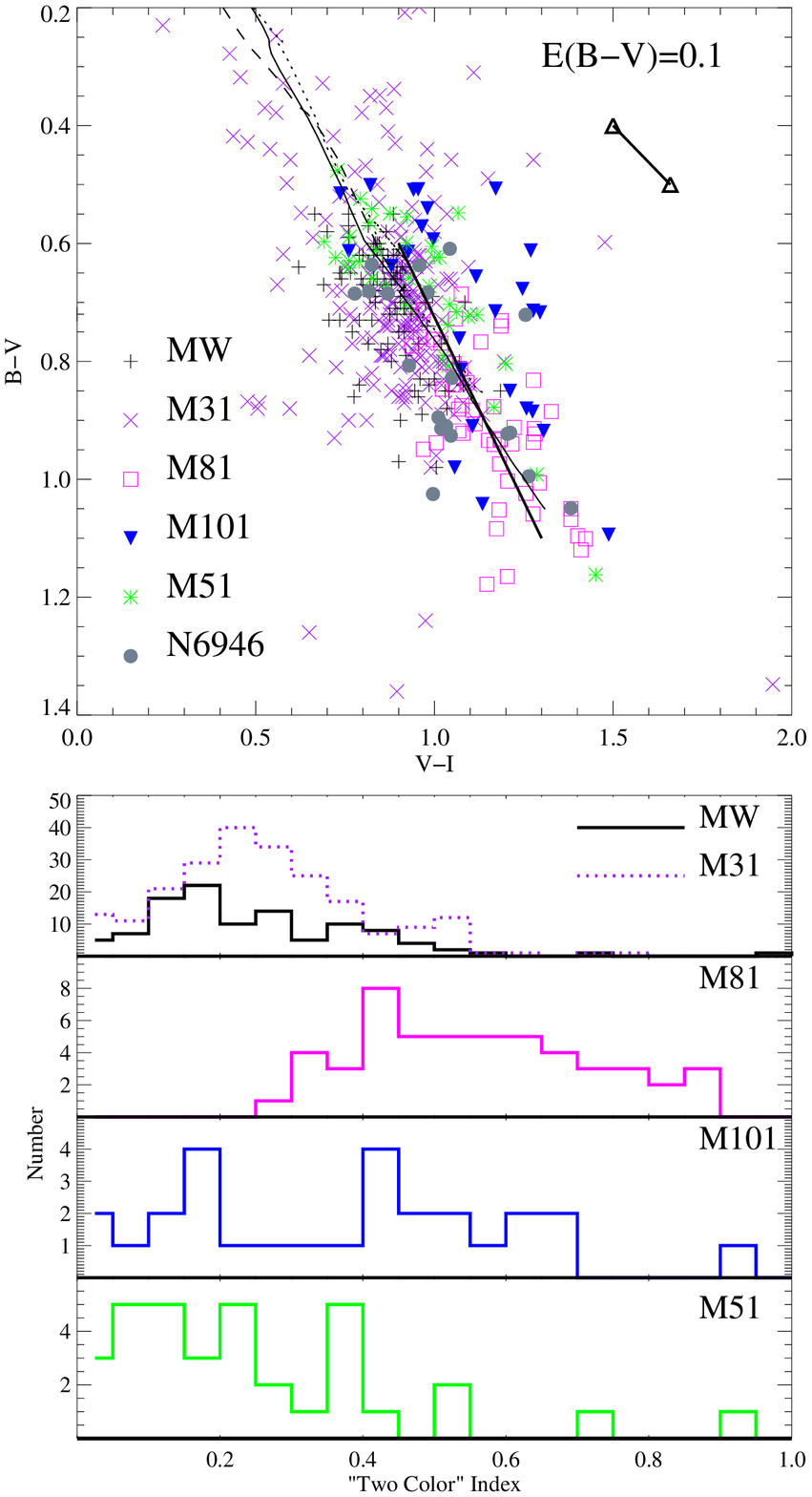}{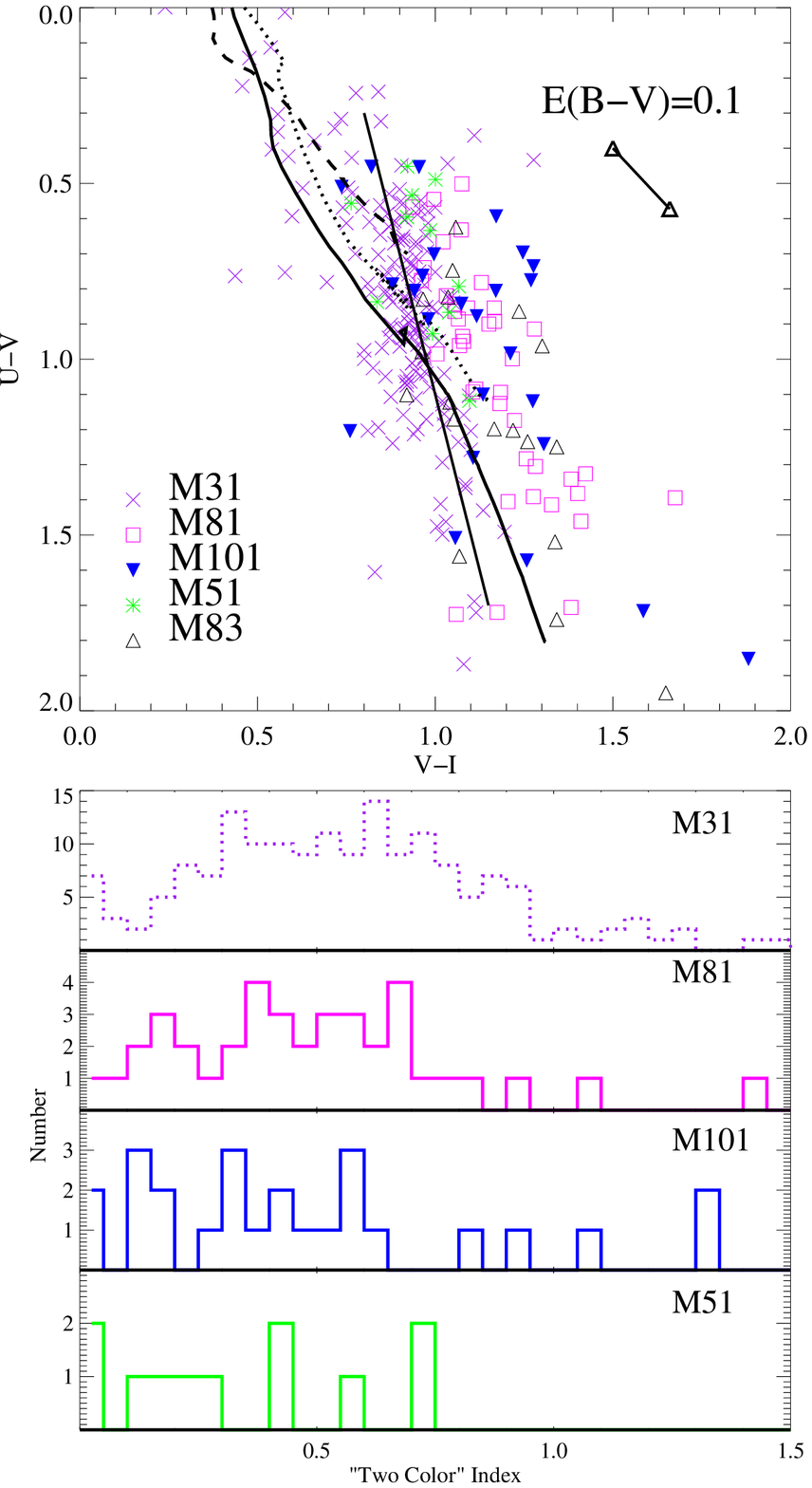}
\caption{{\it Top Panels:} The $(V-I)$ vs. $(B-V)$ (left) and 
$(V-I)$ vs. $(U-V)$ (right) color-color diagrams are shown.
BC00 models and GCs are as in Figure~\ref{2colmetal1}.
{\it Bottom Panels:}  The distribution of cluster positions in color-color
space along the M31 GC locus are shown.  Derivation of these values
are described in $\S4.2$.
\label{2colmetal2}}
\end{figure}

In Figures~\ref{2colmetal2}a,b we show two other color-color
combinations.  In Figure~\ref{2colmetal2}a, which includes most of the
clusters in our sample, we see that four M51 GCs (about 12\%) are
located in the red GC parameter space, while the rest are consistent
with bluer GC colors.  This represents an upper limit to the total
number of metal-rich GCs in our M51 sample (since any intrinsic
reddening would move these objects to bluer colors).

\subsection{Color/Metallicity Distributions of GC systems in spirals}

\begin{deluxetable}{lllcccrrrcccc}             
\tablewidth{0pc}             
\tablecaption{STATISTICS OF COLOR-COLOR DISTRIBUTIONS FOR GC SYSTEMS} 
\scriptsize
\tablehead{             
\colhead{} &
\multicolumn{2}{c}{$(U-B)$ v. $(V-I)$} &
\colhead{} &
\multicolumn{2}{c}{$(U-V)$ v. $(V-I)$} &
\\
\colhead{Galaxy} & 
\colhead{mean} &
\colhead{$\sigma$} & 
\colhead{} & 
\colhead{mean} &
\colhead{$\sigma$} 
}             
\startdata           
Milky Way & 0.38 (0.02) & 0.19 &    & ... & ...   \\
M31 & 0.38 (0.02) & 0.22  &    & 0.56 (0.05) & 0.30     \\
M81 & 0.33 (0.04) & 0.27  &    & 0.50 (0.02) & 0.29   \\
M101 & 0.31 (0.06) & 0.25  &    & 0.48 (0.05)  & 0.37    \\
M51 & 0.17 (0.06) & 0.18 &    & 0.34 (0.07) & 0.24  \\
\tablecomments{Mean and standard deviations are calculated for intrinsic
(dereddened) two color indices.  The technique used to measure this
index is described in $\S4.2$.  The values in parentheses give uncertainties
in the mean, calculated as $\sigma/\sqrt{N}$}
\normalsize
\enddata             
\end{deluxetable}

In this section, we attempt to more fully quantify the metallicity
distributions of spiral GC systems, by using the two color index
developed above.  The intrinsic metallicity distributions for M31 and
Galactic GCs are known to be bimodal.  This translates into an
extended color distribution in the lower panels of
Figures~\ref{2colmetal1} and \ref{2colmetal2}, including both
metal-poor (blue) and metal-rich (red) GCs.  In our cluster samples,
low numbers also compromise our ability to clearly detect bimodality;
because of these small numbers, in general we will refer to
``extended'' metallicity distributions rather than bimodal
distributions.  One way to understand the underlying metallicity
distribution from colors is to compare statistics between systems in
different galaxies.  In Table~8, we compile the mean and standard
deviation for GC two color indices.  These only include clusters with
U band photometry, since we are interested in the intrinsic color
distributions.  We find that the mean and $\sigma$ of the M81 GC
system (0.33 and 0.27 respectively) are very similar to those for M31
(0.38 and 0.22) and the Milky Way (0.38 and 0.19).  M51 has a
significantly lower mean value (0.17) and smaller spread (0.18) than
the other three galaxies.  

Assuming that the M51 clusters studied here are ancient, this is
indicative of lower overall metallicity for the M51 GC system.
Although it is possible that the M51 GC candidates are younger and
therefore more metal-rich, a comparison with stellar evolutionary
models indicates that they would have to be substantially younger than
12~Gyr for this to be true.  For example, in the $U-B$ versus
$V-I$ color-color diagram shown in Figure~\ref{2colmetal1}a, the M51
globular candidates are consistent with the blue, metal-poor M31 and
Milky Way GC colors.  If these were metal-rich the only way for them
to intersect the solar metallicity model (for example), would be if
the reddening was high (with $E_{B-V} \sim0.25-0.3$), and the age
around $10^8$ years.  We consider this possibility unlikely, since the
$E_{B-V}$ distribution for the entire M51 cluster system, including
the youngest objects, is highly peaked at a reddening value much
lower than this.

The ancient clusters in M101 might be expected to be exclusively blue
and metal-poor, since the extremely small bulge in this galaxy makes
it unlikely that a metal-rich population associated with this
component formed.  Although based on small number statistics, the
intrinsic colors of M101 GC candidates appear more similar to those in
M81 than in M51 (mean and $\sigma$ of 0.31 and 0.25 respectively),
consistent with an interpretation that both metal-rich and metal-poor
clusters formed in M101.

A formal test for bimodality is traditionally performed on the color
distributions of elliptical GC systems (e.g., Ashman, Bird, \& Zepf
1994) to better understand their formation histories.  For spiral GC
systems, the complication of variable reddening makes it more
difficult to assess the underlying metallicity distribution based only
on integrated colors.  
When using individual colors to test for
bimodality in the M31 GC system, Barmby et~al.\  (2000) found that only
two optical colors, $(U-V)_0$ and $(U-R)_0$ showed evidence for
bimodality at the 95\% confidence level.  They found that photometric
errors are likely large enough to mask any color separation in most
single color distributions for GCs in Andromeda and the Galaxy.
Previously, we found no evidence for bimodality in the $(B-V)$,
$(B-I)$, or $(V-I)$ color distributions of M81 GCs (Chandar,
Tsvetanov, \& Ford 2001).  Rather than repeating the test for
bimodality on single color distributions, here we use our two color
statistic to probe underlying metallicity distributions.  We restrict
our samples to objects which have U band photometry, since this filter
is crucial according to the Barmby et~al.\  2000 results, and also
allows us to determine intrinsic (dereddened) cluster colors.  

First, we tested the color-color distributions of M31 GCs using the
KMM algorithm (McLachlan \& Basford 1988; Ashman et~al.\  1994).  As
input to the KMM algorithm, we used the two color index values (as
derived in the previous section), an initial mean and dispersion for
the two Gaussian groups to be fit (the final solution is not very
sensitive to these starting points unless there are many outliers),
and the relative proportion of objects in each group.  The p-value
returned by KMM for a given distribution measures the statistical
significance of the improvement in the fit when going from a single
gaussian to (in this case) two gaussians.  For M31 and the Milky Way,
the hypothesis of a unimodal distribution in our $U-B$ vs. $V-I$ two
color index space was rejected at the $>95$\% confidence level.
Similarly, when we tested the M81 distributions including U band
photometry, a unimodal distribution is rejected at the $\sim90$\%
confidence level, although a minimum of 50 data points should be used
to obtain a reliable result.  Peaks near values of 0.20 and 0.45 were
found by the KMM algorithm for the Milky Way and M31 GC systems.
If we estimate the relative fraction of metal-poor to metal-rich GCs
in our M81 sample by making a simple cut at a two color index of
0.325, we find that roughly 60\% of the M81 GCs with UBVI photometry
are consistent with their metal-poor Galactic and M31 counterparts.
However, we caution that the location of the archival fields in M81
bias our sample against metal-rich bulge globulars. 

In conclusion, we find that the dereddened color distributions of M81
and M101 GCs are consistent with an interpretation of an extended
metallicity distribution similar to that found in the Milky Way and
M31 GC systems, whereas in M51 most ancient clusters appear to be
metal-poor.

\subsection{Size Distributions}

The structures of GCs yield information concerning their formation and
the environmental influence of the host galaxy.  There is some
evidence for differences in the mean structural properties of clusters
between galaxies.  For example, GCs in the LMC are more flattened on
average than their counterparts in the Milky Way (e.g., Geisler \&
Hodge; Frenk \& Fall 1982).  Much more evidence points to a size
difference between red, metal-rich and blue, metal-poor GC
subpopulations {\it within} galaxies, with red clusters being
systematically more compact (for results in early-type galaxies, see
e.g., Kundu \& Whitmore 1998; in Andromeda, see Barmby, Holland, \&
Huchra 2002).

Intrinsic sizes for GCs were measured (from V band images)
using the ISHAPE routine.  A detailed description of the code is given
in Larsen (1999), along with the results of extensive performance
testing.  Essentially, ISHAPE measures intrinsic object sizes by
adopting an analytic model of the source and convolving this model
with a (user-supplied) PSF, and then adjusting
the shape parameters until the best match is obtained.  King model
profiles with concentration parameters of $c=30$ were convolved with a
PSF, and fit individually to each object.  ISHAPE estimates the FWHM
of each cluster (in pixels), which was then converted to the
half-light (effective), $r_{eff}$, by multiplying the FWHM 
by a factor of 1.48,
as described in the ISHAPE manual.

To measure cluster sizes in galaxies beyond the Local Group, it is
important to have a good characterization of the PSF.  We selected M51
(the most distant galaxy in our sample, and thus the most likely to
present difficulties in measuring sizes) to test two different
techniques: 1) using hand-selected, relatively isolated stars, and 2)
a theoretical PSF created from the TinyTim routine (Krist 1995).  We
found that the size estimates from ISHAPE using these two PSFs
differed by less than 20\%.  Final size measurements for M81, M83,
NGC~6946, and M51 were made using a TinyTim PSF,
since this is easily reproducible.  One PSF was generated for the PC
CCD, and one for the WF CCDs.  Sizes measured independently for eight
M81 clusters located in overlapping \textit{HST} images agreed to better than
10\%.  Because these objects are located in different portions of CCDs
in the two observations, the level of agreement indicates that using a
single PSF for each CCD is sufficient.  However, we caution that
focusing issues, and distortions could cause the intrinsic PSF to be
slightly broader for clusters located near the edge of a CCD.

For M101 we implemented a different procedure, since our images were
drizzled together.  Details of our measurement technique and testing
will be presented in an upcoming paper (Converse, Chandar, \& Whitmore
2004, in prep).  Briefly, we created a TINYTIM PSF at the original
location of each M101 cluster, and then drizzled this PSF by itself
exactly as was done for the data.  A comparison of bright clusters
with sizes measured from both drizzled and non-drizzled M101 images
showed excellent agreement.

One very interesting result concerning the size distribution of star
cluster systems is the example of NGC~1023.  Larsen \& Brodie (2000)
found an excess of faint clusters in this lenticular galaxy, relative
to the expected turnover in the luminosity distribution.  These faint,
ancient clusters differ in (at least three ways) from their more
luminous counterparts: {\it i)} they are more diffuse, {\it ii)} they
are more metal rich on average, and {\it iii)} they appear to have
disk-like kinematics, with a strong rotation signature (Brodie \&
Larsen 2002).  When NGC~1023 clusters are separated by size at
$\sim7$~pc, the luminosity function for the compact clusters has a
turnover near $M_V\sim-7.5$, while the excess faint clusters continue
in power-law fashion to the detection limit.

\begin{figure}
\epsscale{0.4}
\plotone{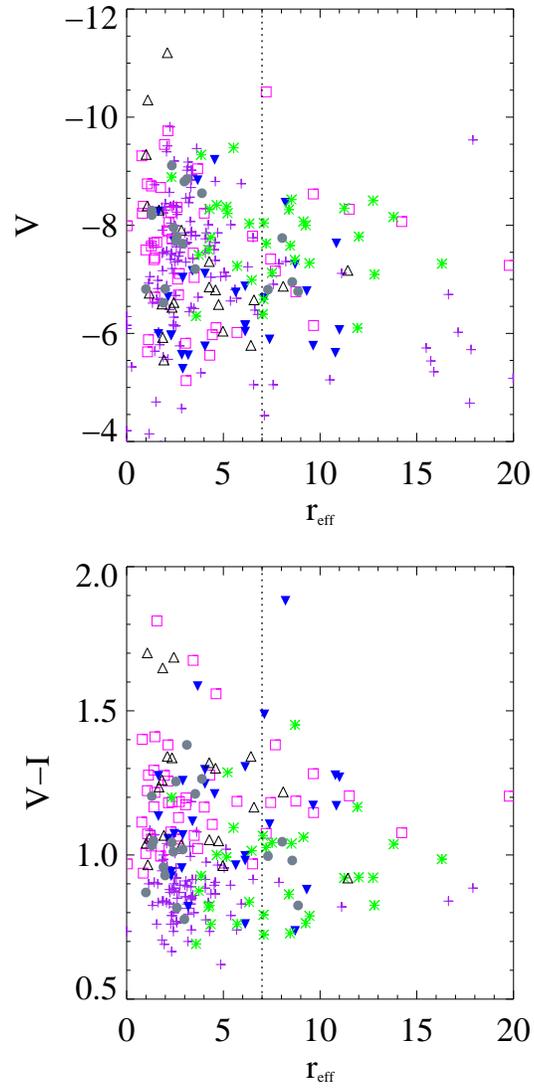}
\caption{Measured effective radii for our GCs.  Purple crosses
are the half-mass radii of the Galactic system.  The dashed line
at 7 pc is that used by Larsen \& Brodie (2000) to separate ``faint fuzzies''
from more compact clusters in NGC~1023.
\label{size}}
\end{figure}

Figure~\ref{size} shows effective radii for ancient cluster candidates
in all five galaxies as a function of luminosity and color.  For
comparison, we have added the half-mass radii of the Galactic GC
system (Harris 1996).  
The dashed line marks an effective radius of 7 pc.
In the Galaxy only $\sim13$\% of GCs are more extended than 7 pc.
Our GC samples have extended cluster fractions range from 10\% (M83)
to 59\% (M51).  Formally, a Kolmogorov-Smirnoff (K-S) test finds that
the size distributions in the Milky Way and M51 GC systems differ at a
confidence level $>99$\%. The results for the other cluster systems
are less conclusive, but show a much higher probability (up
to 85\%) that they are drawn from the same parent distribution as the
Milky Way GCs.

In the previous section, we reported the detection of a number of
faint ancient cluster candidates in M101 and NGC~6946.  These have a
similar {\it luminosity} distribution to their faint counterparts in
NGC~1023.  However, the {\it size} distributions for the faint M101
and NGC~6946 clusters discovered in this work are similar to those of
the more luminous GCs in their respective parent galaxies.

Could the observed difference of the M51 GC system simply be due to
 observational bias? 
The \textit{HST} imaging is sufficient to select compact clusters in M51.  In
 fact our catalog of young M51 clusters has a large fraction of
 compact objects (with $r_{eff}\sim1.5-2$~pc).
Finally, we note that the ancient cluster candidates in M51 are almost
 exclusively blue, while those in M81 are preferentially red.
 Although the physical reason for this difference isn't clear, these
 results are broadly consistent with previous observations of a
 size-metallicity trend, since the M51 GCs appear significantly
 bluer (and more diffuse) than their redder, more compact M81
 counterparts.

\subsection{Numbers and Specific Frequencies of Globular Clusters in Spirals}

\subsubsection{Total Number of Globular Clusters}

Our final globular cluster samples include 47, 21, 19, 29, and 34
objects for M81, M83, NGC~6946, M101, and M51 respectively.  Given the
very low fractional contamination estimated in $\S3.5.1$, our survey
provides unambiguous evidence that globular cluster systems exist in
each of our target galaxies, despite the fact that M101, M83, and
NGC~6946 are all of Hubble type Sc or later.  In this section, we
attempt to determine the total number of GCs associated with each host
galaxy.  We use the following basic recipe:

\begin{itemize}
\item 
We assume that the GC luminosity function turnover occurs at
$M_V=-7.4$.  The total number of GCs is defined as twice the number of
GCs brighter than the turnover magnitude of the GCLF, where the GCLF
is assumed to be a Gaussian (in magnitude units).

\item The results of the artificial cluster experiments described in
$\S3.5.2$ are used to estimate the completeness of our GC samples.  We
determined an incompleteness fraction for each GC based on its V band
luminosity.
Each bin in the luminosity function was divided by the average
completeness fraction of all objects in that bin to produce a
completeness corrected value.  Because we don't track local background
levels for the clusters, our completeness corrections are ``averaged''
over a range of environments, which we assume to be representative
for the entire GC population.

\item We then sum up the number of (completeness corrected) globular
clusters to the expected turnover, and multiply this value by
a factor of 2, to account for the faint half of the distribution.

\item Finally, we correct for the (limited) spatial coverage of the
galaxy in our survey.  Our technique for estimating the correction or
scale factor for the galaxies studied in this work is described below.
Note however, that our technique to correct for this should not be
considered a replacement for imaging data which has broader coverage.

\end{itemize}

For spirals, two techniques have generally been used to make a
correction for limited galaxy coverage.  Larsen et~al.\  (2001)
constructed radial distribution functions of GCs.  The caveat to this
technique is that it requires a large population of GCs to avoid
complications resulting from small number statistics.  Kissler-Patig
et~al.\  (1999) and Goudfrooij et~al.\   (2003) correct for spatial
coverage by making a direct comparison with the GC locations in the
Milky Way.  This technique makes the implicit assumption that GC
systems in external spirals have a similar spatial distribution as the
GC system in our Galaxy.  Goudfrooij et~al.\  (2003) find that the total
number of GCs estimated for NGC~4594 using both methods described
above gives consistent results.  The estimated GC population in
NGC~7814 is also similar between the two techniques (Goudfrooij et
al. estimate $106\pm28$ total GCs using the Galaxy-comparison
technique, and Rhode \& Zepf (2003) find $140-190$ GCs by fitting the
radial profile of the GC system).  Because we do not have sufficient
numbers of clusters or radial coverage to use the former technique, we
devised a procedure similar to the latter.  We use data from the Milky
Way GC system compiled in the McMaster catalog (Harris 1996), which
contains 150 GCs.  However, we adopt $N_{MW}=160\pm20$ (van den Bergh
1999) as the total number of Galactic GCs.  The undetected MW GCs are
assumed to lie behind the Galactic bulge, and the locations of these
``missing'' clusters was synthesized by reflecting 10 known clusters
within 2.0 kpc of the bulge in the projected ``Y-Z'' plane.  In the
following discussion of Galactic GC locations, X points toward the
Galactic center, Y points in the direction of Galactic rotation, and Z
toward the North Galactic Pole.  There are essentially two different
orientations which can be considered for external galaxies viewed
face-on, corresponding to clusters with $+$ or $-$ Z locations (which
side of the disk is observed).  We created a mask defined by our
spatial coverage of each galaxy, and then applied this mask to both
face-on presentations of the Milky Way (i.e., the $\pm\mbox{Z}$
locations projected onto the X-Y plane of the Galactic disk).  By
calculating the fraction of the total GC system observable in each
mask, we were able to determine a ``scale factor'' for the incomplete
spatial coverage of our observations (taken to be the average from the
two orientations): $S_{complete}=N_{MW}/N_{mask}$, where $N_{mask}$ is
the number of GCs detected in the mask, and $N_{MW}$ is the total
number of GCs in the Milky Way system.  This technique makes the
explicit assumption that GCs are found predominantly associated with
bulges and/or halos of galaxies.  If instead globular clusters in a
given galaxy are associated with a thin disk,
and can be seen above the
dust layer from either side, then our technique will overestimate the
number of clusters.

Our technique to derive the total number of GCs can be written as:

\begin{equation}
N_{GC} = 2 \times S_{complete} \sum_{i=V_{min}}^{V_{turnover}} \frac{1}{\langle f_{i} \rangle}~N_{V,i}
\end{equation}

where $S_{complete}$ is the scale factor used to correct for the
limited spatial coverage of the observations, the luminosity fuction
is summed from the brightest magnitude bin $V_{min}$ to the magnitude
bin covering the GCLF turnover; $\frac{1}{\langle f_{i} \rangle}$
represents the average fractional completeness for GCs in a given
luminosity bin, and $N_{V,i}$ is the number of clusters observed in
a given luminosity bin.

As discussed in $\S4.1$, M101 and NGC~6946 do not appear to have a
typical log normal GC luminosity distribution in magnitude space.  The
observations probe these galaxies deeply enough that we can push
$\sim1.5$ magnitudes beyond the expected turnover in the GC luminosity
function.  Having done this, we find an excess of faint, red, resolved
clusters, a population which clearly does not exist in M81.  Whether
these objects are true ancient GCs formed early in the universe, or
whether they have ages of a few billion years remains to be
determined.  For the purposes of discussing the total number of GCs in
the three latest-type galaxies, we simply sum up the (completess
corrected) clusters to the expected turnover ($-7.4$), and follow
equation 1, thus {\it explicitly excluding} this ``excess'' faint
population.  In M101 and M83 if we sum up the (completeness corrected)
GC sample to $M_V=-6$, the total number of clusters increases by a
factor $\sim1.8$.  Because of M81's proximity, we can reach almost the
entire expected GC population.  For this galaxy, we tried two
approaches.  First, we used the bright half of the GC luminosity
function as representative of the faint portion, and second we summed
the entire completeness corrected cluster population.  Both techniques
result in similar total GC populations for M81 (430 vs. 450), and we
retain the numbers based on the second method.  

Uncertainties in the total number of GCs are dominated by the
correction for limited coverage.  The distribution of Galactic GCs
becomes stochastic as one moves away from the galaxy center, since
they are not evenly projected in the X-Y plane at larger
galactocentric distances.  We attempted to place limits on the upper
and lower fractional coverage by considering the full range of
clusters covered if a given outer WFPC2 pointing was located at the
same physical distance from the galaxy center, but at a different
location.  Additionally, uncertainties in the total number of clusters
based on the unavailability of the $U$ filter for some fields were
considered.  The range in fractional coverage and likely contamination
fraction were translated into uncertainties in the total number of GCs
derived for each galaxy.

The calculated total GC numbers residing in each target galaxy and
associated uncertainties are recorded in column 6 of Table~9.
Previously, we estimated the M81 GC population to be $211\pm29$
(Chandar, Tsvetanov, \& Ford
2001).  There were two weaknesses in our previous technique: 1) we
didn't explicitly make completeness corrections, and 2) our
measurements hinged on fitting the radial profile of the GC system,
which has very large uncertainties.  
The new results presented here supercede our previous numbers.

We make an independent consistency check on the number of GCs derived
in M81 as follows.  Using a single WFPC2 V band image, Davidge \&
Courteau (1999) estimated that $45\pm12$ GCs brighter than $M_V=-7$
reside within the central 2 kpc of M81.  Using our technique above,
the MW GC system, when projected onto M81 (and accounting for
distance, inclination and position angle), has $\sim20$\% of the
visible population (on a given side of the disk) in the same area.
This implies $48\pm15$ GCs based on our estimated GC population, in
good agreement with the Davidge \& Courteau observations.

\begin{deluxetable}{lllcccrrrcccc}             
\tablewidth{0pc}             
\tablecaption{NUMBERS OF GCs AND SPECIFIC FREQUENCIES FOR SAMPLE GALAXIES } 
\tablehead{             
\colhead{Galaxy} & 
\colhead{B/T\tablenotemark{a}} &
\colhead{$M_V^{0}$\tablenotemark{b}} &
\colhead{$M_V$} & 
\colhead{GC}  & 
\colhead{GC} & 
\colhead{$S_N$\tablenotemark{c}} &
\colhead{T\tablenotemark{c}} & 
\\
\colhead{} & 
\colhead{} &
\colhead{} &
\colhead{bulge} &
\colhead{(det)} &
\colhead{(total)} &
\colhead{total} &
\colhead{total} &
}             
\startdata           
M81 & 0.46 & $-21.63$ & $-20.60$ & 47 & $450\pm145$ & $1.0\pm0.3$ & $1.9\pm0.5$\\
M83 & 0.05 & $-21.01$ & $-18.51^{+0.75}_{-0.44}$  & 21 & $150\pm20\phn$  & $0.6\pm0.1$ & $1.4\pm0.2$\\
NGC~6946 & 0.02 & $-21.46$ &  $-18.21_{-0.75}$  & 19 & $90\pm40$ & $0.2\pm0.1$ & $0.7\pm0.3$\\
M101 & 0.04 & $-21.42$ & $-18.17_{-0.75}$ & 29 & $150\pm40\phn$ & $0.4\pm0.1$ & $1.2\pm0.3$\\
M51 & 0.42 & $-21.68$ &  $-20.73$  & 34 & $220\pm45\phn$ & $0.5\pm0.1$ & $1.1\pm0.2$\\
\tablecomments{
The derivation of the total number of clusters is given in $\S4.5.1$, and
explicitly excludes the faint, excess clusters in M101 and NGC~6946.}
\tablenotetext{a}{The Bulge/Total ratios are based on bulge/disk 
decompositions.  We adopt the values derived from the Baggett et~al.\  (1998)
fits for M81 and M51, and from our fits to 2MASS K-band images for M83,
NGC~6946, and M101.  See $\S4.5.2$ for details}
\tablenotetext{b}{From Lyon Extragalactic Database (LEDA; http://leda.univ-lyon1.fr/).  The absolute V magnitudes have been corrected for both foreground and internal extinction.}
\tablenotetext{c}{The given errors only reflect uncertainties in the total
number of estimated GCs.  If the total galaxy magnitudes are uncertain
by $\pm0.2$~mags, this would add roughly $\pm0.03$ and $\pm0.1$ in 
uncertainty to the $S_N$ and T values respectively.}
\normalsize
\enddata             
\end{deluxetable}

\subsubsection{Bulge Luminosities}

One of the goals of this work is to study the red metal-rich
and blue metal-poor GC populations separately.  In the Milky Way,
there has been much recent evidence to support the view that inner,
metal-rich GCs in spirals are associated with the Galactic bulge
rather than the disk (Minniti 1995; Cote 1999).  
In order to test this concept, we need information on the relative
contributions of the bulge and disk.

Because GC specific frequencies are traditionally normalized to the
absolute V magnitude of the host galaxy, and by extension the
estimated V magnitude for the bulge, it is preferable to use
bulge/disk decompositions measured from a similar passband, such as
found in Baggett, Baggett, \& Anderson (1998).  As a check on the
Baggett et~al.\  (1998) results, we downloaded K band 2MASS images of
our target galaxies and performed our own decompositions.  It has been
suggested that the near infrared is the ideal wavelength regime to
study the stellar populations which make the dominant mass
distribution in a galaxy (e.g., Rix \& Rieke 1993).  There are two
main reasons for this.  First, the extinction is lower by a factor of
10 between the B and K bandpasses, and second, the emission of old
stellar populations peaks in the NIR.  As bulge/disk decompositions
are not the main goal of this work, we only briefly describe our
techniques.  We used the IRAF task ELLIPSE to create the surface
brightness profiles, and then compared the results of fitting a double
exponential vs. a de Vaucouleurs profile plus an exponential.  These
resulted in K band measurements of bulge and disk effective radii.  To
determine the K band surface brightness at these effective radii, we
use the extended source 2MASS on-line catalog
(http://pegasus.phast.umass.edu), where K band surface brightnesses at
our best fit bulge and disk effective radii were read off from the
available profiles.  These values were then used to estimate the
bulge-to-total (B/T) K-band luminosity ratios.  If the Baggett et
al. (1998) values differed significantly from our decompositions, we
use the B/T values implied by our K-band fits to determine the bulge V
band magnitude.  In general, we find that M101, M83, and NGC~6946 (the
three latest type galaxies) are best fit by a double exponential
profile (although the double nucleus in M83 makes the results of our
fit uncertain for this galaxy), and the results of our fits are
adopted.  The earlier type galaxies M51 and M81 are best fit by de
Vaucouleurs bulge profile, plus an exponential disk, and we obtain
results quite similar to those of Baggett et~al.\  (1998).  Column 2 of
Table 9 lists the adopted B/T values, column 3 gives the total V band
galaxy luminosity, and column 4 gives the associated bulge
luminosities.

\subsubsection{Specific Frequencies}

The total number of GCs can be normalized by galaxy luminosity or mass
to facilitate comparison with other GC systems.  The specific
frequency, $S_N$ (as defined by Harris \& van den Bergh 1981)
$S_N\equiv N_{GC}\times10^{+0.4(M_V+15)}$, is the number of GCs
normalized by the total galaxy luminosity.  Because different galaxy
types are dominated by different stellar populations, normalizing by
galaxy mass may give a more consistent comparison between galaxies
with different star formation histories.  Zepf \& Ashman (1993) define
$T\equiv \frac{N_{GC}}{M_G/10^9~M_{\odot}}$, where $M/L_v$ ratios of
6.1 (Sab-Sb), 5.0 (Sbc-Sc), and 4.0 (Scd-Sd) are used to convert
galaxy luminosity to mass.  The total T and $S_N$ values for our
galaxy sample are presented in columns 7 and 8 of Table~9.

\section{DISCUSSION}

\subsection{What is the nature of the faint clusters in late-type
spirals?}

As discussed in $\S4.1$, NGC~6946 and M101 have formed a number of
faint clusters which have integrated colors indistinguishable from
globular clusters.  Note however, that the U band photometric
uncertainties are quite large for many of these.  The presence of
faint, ancient clusters has a significant influence on the observed
luminosity distribution for our GC candidates; rather than turning
over as expected, the clusters apparently follow a powerlaw
distribution to the detection limit.  Earlier, we found a similar
situation in the late-type Local Group spiral M33 (Chandar, Bianchi,
\& Ford 2001).  Qualitatively, the luminosity distribution of the
faint GC candidates is similar to that found by Larsen \& Brodie
(2000) for faint, red, extended clusters in NGC~1023.  Follow-up
spectroscopy has established that the ``faint fuzzies'' in NGC~1023
rotate with the disk, unlike the more luminous, compact GCs, which
have kinematics expected for halo/bulge objects.  These faint diffuse
clusters are ancient ($\geq7$ Gyr), and relatively metal rich (Brodie
\& Larsen 2002).  The faint M101 and NGC~6946 objects differ from
``faint fuzzies'' in that they are compact ($\S4.4$), with sizes
indistinguishable from the more luminous GC candidates in M101, and
they are not preferentially red (metal-rich).

\begin{figure}
\epsscale{0.6}
\plotone{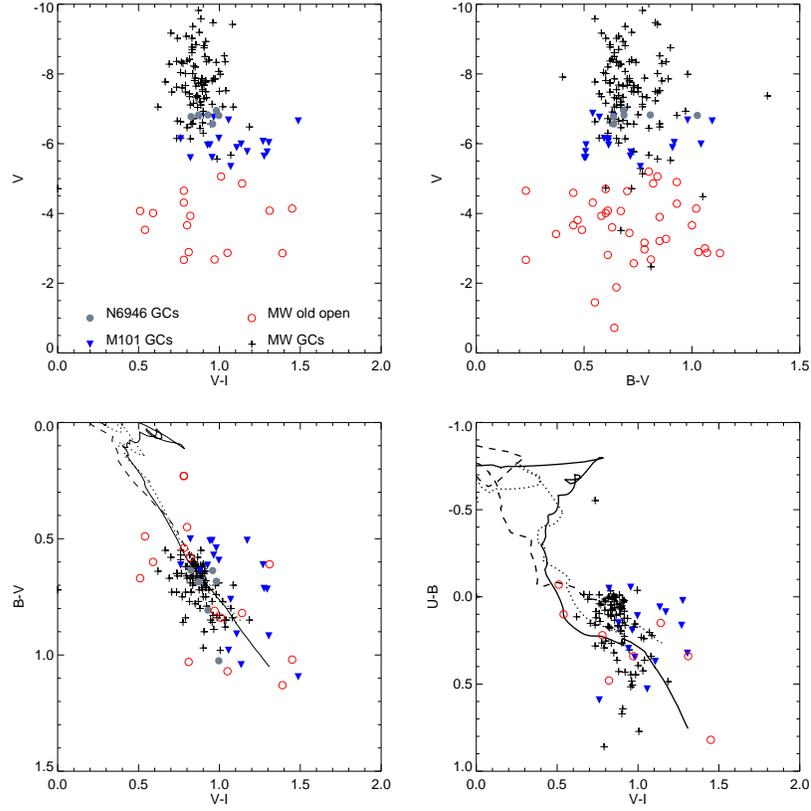}
\caption{The upper panels shows color-magnitude diagrams for
``faint'' M101 (blue triangles) and NGC~6946 (gray filled circles)
globular cluster candidates (defined as $M_V\geq-7$).  These are compared
with the absolute luminosities and dereddened colors of Galactic globular
clusters (crosses) and old ($>9.0$ log yrs) open clusters (red open circles).
The lower panels shows color-color distributions for all four cluster
populations.
\label{faintcl}}
\end{figure}

What is the nature of these faint clusters?  Are they truly ancient,
low mass clusters as suggested by their luminosities and colors?
Because they are resolved, they cannot be individual stars, and their
round morphology and location in the inner portions of the host galaxy
(where almost no background galaxies are seen) is incompatible with a
population of background galaxies.
Are these analagous to the old open
clusters in the Milky Way; ancient ($1-9$ Gyr), less massive (few
$\times10^3~M_{\odot}$) clusters residing in the (thin) Galactic disk?
In Figure~\ref{faintcl} we plot the color magnitude (upper panels) and
color-color diagrams (lower panels) of old ($\geq 9.0$ log yrs) open
Galactic clusters with available integrated photometry (Lata et
al. 2002; Mermilliod \& Paunzen 2003), and also include M101 and
NGC~6946 clusters fainter than $M_V=-7$.  For comparison, we also plot
the Galactic GC population.  Figure~\ref{faintcl} illustrates two
points. First, the old open clusters in the Galaxy appear fainter than
the objects discovered in M101.  However, current surveys of
disk clusters are probably very incomplete due to large extinction
in the disk.  Thus it is likely that additional old open clusters
exist, and these could overlap in luminosity with the objects
presented here.  The faint M101 clusters are defined so they lie in a
luminosity range where Galactic GCs are falling off, and observational
detection limits prevent us from observing old clusters as faint as
known Galactic old open clusters.  However, we cannot rule out that
deeper photometry might reveal populations which overlap with old open
cluster luminosities.
The second point is that there is significant overlap in
the colors of GCs and old open clusters (which are many Gyr younger).
Partially, this is due to the age-metallicity degeneracy, since the
somewhat more metal rich open clusters have slightly redder colors for
their age than do the more metal-poor, older GCs.  
M101 cluster colors are consistent with formal age estimates $\gea
3$~Gyr based on comparison with sub-solar BC00 models.  Follow-up
spectroscopy is needed to establish whether faint M101 and NGC~6946
clusters are younger than their more luminous counterparts and whether
they reside in a disk.

While we cannot rule out that the faint M101/NGC~6946 clusters are
counterparts to old Galactic {\it disk} clusters, a second possibility
is that the faint red clusters are the low mass extension of the
initial {\it halo} GC population, which were somehow able to survive
internal and external dynamical evolution for a Hubble time.  To
explore this possibility, we compared available velocities for GCs in
M33 fainter than $M_V=-7$, with their more
luminous counterparts.  Although there are only 17 such faint M33 GCs
with available velocities, these have (halo) kinematics
indistinguishable from more luminous M33 GCs (velocities taken from
Chandar et~al.\  2002).  Therefore in M33, regardless of the exact
origin of ancient halo clusters, it appears that a larger fraction of
lower massive clusters have been able to survive destruction over a
Hubble time than found in the Milky Way or M31.  The fact that these
excess faint anceint clusters have only been seen in the latest-type
spirals, where potentially the dynamical conditions are different from
those in other well studied but earlier type galaxies, may provide
clues to their survival.

\subsection{Is There a Case for ``Universal'' halo globular cluster systems?}

Goudfrooij et~al.\  (2003) studied edge-on spiral galaxies of different
Hubble types.  Based on their result that spirals with $B/T\lea0.3$
(i.e.  Hubble type later than Sb) have very similar GC specific
frequencies ($S_N = 0.55\pm0.25$), they suggest that this population
represents a ``universal'', old halo population which is present
around each galaxy.  Rhode \& Zepf (2003) found a similar
mass-normalized total number of metal-poor (presumably halo) GCs in
the Sab spiral NGC~7814, as well as in the modest luminosity
elliptical galaxy NGC~3379.

This is an important concept to test, considering the variety observed
among the four most massive Local Group galaxies.  
While M33 has mostly halo GCs, in the LMC all ancient clusters reside
in the (thin) disk (Freeman, Illingworth, \& Oemler 1983).  Thus, it
is not clear whether {\it all} disk galaxies form halo systems of
ancient clusters.
In columns 7 and 8 of Table~9 we have compiled the estimated $S_N$ and
T values for the galaxies presented in this work.  We find that for
all sample galaxies of type later than Sb (this only excludes M81, the
earliest and most bulge dominated galaxy in our sample), the $S_N$ and
T values are very similar, around 0.5 and $1.0-1.2$ respectively.
Note that for M101 and NGC~6946, we have disregarded the population of
faint, red clusters in deriving the estimated total numbers of GCs in
these host galaxies.  
In M81, the observed $\sim60$\% blue fraction in our sample is similar
to that found in the Milky Way and M31 GC systems.
This results in 
$S_{N,blue}\sim0.6$ and $T_{\mbox{blue}}\sim1.1-1.2$, in the same range as we
find for the later type spirals in our sample.

We conclude that all galaxies studied in this work, both early- and
late- type, support the concept that (massive) spirals have formed a
similar number of mass-normalized total GCs, with $S_N\sim0.5$ and
$\mbox{T}\sim1.3\pm0.2$.  Since later-type galaxies are dominated in
mass by their halos, this suggests that a ``universal'' population of
halo GCs may have formed associated with all (massive) galaxies.  This
statement is predicated on the assumption that if the faint (low mass)
clusters discovered in our Sc and later-type galaxies
belong to the original halo GC population, their presence is
the result of a difference in {\it destruction} timescales rather
than formation mechanisms.

\begin{deluxetable}{lllccrccccccc}             
\tablewidth{0pc}             
\tablecaption{``SCORESHEET'' FOR GLOBULAR CLUSTER SYSTEMS IN INDIVIDUAL SPIRAL GALAXIES} 
\scriptsize
\tablehead{             
\colhead{Galaxy} &
\colhead{$M_V$} &
\colhead{Type} & 
\colhead{$S_N$} & 
\colhead{T}  & 
\colhead{$M_V$} & 
\colhead{$B/R$} & 
\colhead{univ} & 
\colhead{univ} & 
\colhead{disk} &
\colhead{$r_\mathrm{eff}$} & 
\colhead{} 
\\
\colhead{} & 
\colhead{} &
\colhead{} &
\colhead{} &
\colhead{} &
\colhead{peak} &
\colhead{frac} &
\colhead{halo?} &
\colhead{bulge?} &
\colhead{subpop} &
\colhead{} &
\colhead{Refs}
}             
\startdata           
1. MW & $-21.3$ & Sbc   & 0.6$\pm$0.1 & 1.3$\pm$0.4 & $-7.4$  & 0.70 & Y & Y & Y & Y & 1\\
2. M31 & $-21.8$ & Sb   &0.9$\pm$0.2 & 1.6$\pm$0.4 & $-7.4$  & 0.66 & Y & Y & Y & Y & 2\\
3. M33 & $-19.4$ & Scd   & 0.6$\pm$0.1\rlap{\tablenotemark{a}} & 1.6$\pm$0.3\tablenotemark{a} & $\lea\!-7.0$ & \nodata &  Y & \nodata  & Y\rlap{:} & Y & 3 \\
4. NGC~55 & $-19.5$ &  Sm  & 0.3$\pm$0.2\rlap{:}  & 1.6$\pm$0.3 & \nodata & \nodata & \nodata & \nodata & \nodata & \nodata & 4 \\
5. NGC~253 & $-20.2$ &  Sc  & 0.5$\pm$0.3\rlap{:} & 1.1$\pm$0.5\rlap{:} & \nodata & \nodata & \nodata & \nodata & \nodata & \nodata  & 4,5\\
6. M81 & $-21.63$ & Sab   & 1.0$\pm$0.3 & 1.9$\pm$0.5 & $-7.5$ & 0.6\rlap{:} & Y & Y\rlap{:} & Y  & Y  & 6 \\
7. NGC~6946 & $-21.46$ & Sc & 0.2$\pm$0.1 & 0.7$\pm$0.3 & $\lea\!-7.0$ & \nodata & Y\rlap{:} & \nodata & \nodata & Y & 6\\
8. M83 & $-21.01$ & Sc & 0.6$\pm$0.1   & 1.4$\pm$0.2 & $\sim\!-7.5$ & \nodata & Y & \nodata &  \nodata & Y & 6\\
9. M101 & $-21.42$ & Scd   & 0.4$\pm$0.1 & 1.2$\pm$0.3 & $\lea-\!6.0$ & \nodata & Y & \nodata & \nodata & Y & 6\\
10. M51 & $-21.68$ & Sbc   & 0.5$\pm$0.1\rlap{0} & 1.1$\pm$0.2 & $<\!-8.0$  & 1.0 & Y & N & \nodata  & N & 6\\
11. NGC~4594 & $-22.20$  & Sa & 1.7$\pm$0.6 & 3.6$\pm$1.1 & \nodata & \nodata & Y & Y & \nodata & \nodata & 7\\
12. NGC~3628 & $-21.03$  & Sb & 0.6$\pm$0.1\rlap{\tablenotemark{b}} & 0.7$\pm$0.2\rlap{\tablenotemark{b}} &  & \nodata & Y & Y & \nodata & \nodata  & 7\\
13. NGC~4565 & $-21.48$  & Sb & 0.6$\pm$0.2 & 1.0$\pm$0.3 & \nodata & \nodata & Y & \nodata  & \nodata & \nodata & 7\\
14a. NGC~7814 & $-20.46$  & Sab & 0.7$\pm$0.2 & 1.5$\pm$0.5   & \nodata & \nodata & Y & N  & \nodata & \nodata   & 7\\
14b. NGC~7814 & $-20.46$  & Sab & 1.3$\pm$0.4 & 2.2$\pm$0.8   & \nodata & \nodata & Y & Y\rlap{:}  & \nodata & \nodata   & 5\\
15. NGC~4013 & $-20.83$  & Sb & 1.1$\pm$0.3 & 2.2$\pm$0.7 & \nodata & \nodata & Y & Y & \nodata & \nodata & 7\\
16. NGC~4517 & $-21.64$  & Sc & 0.6$\pm$0.2 & 1.4$\pm$0.5 & \nodata & \nodata & Y & \nodata & \nodata & & 7\\
17. IC~5176 & $-21.09$  & Sbc  & 0.5$\pm$0.1 & 1.1$\pm$0.3 & \nodata & \nodata & Y & N\rlap{:} & \nodata & \nodata & 7 \\
\tablerefs{
(1) Ashman \& Zepf 1998
(2) Battistini et~al.\  1993
(3) Chandar et~al.\  2001
(4) Beasley \& Sharples 2000
(5) Rhode 2003
(6) this work
(7) Goudfrooij et~al.\  2003
}
\tablenotetext{a}{Chandar et~al.\  (2001) estimated the total GC population in
M33 as $75\pm14$.  However, this includes clusters fainter than the turnover
magnitude of $M_V\sim-7.4$ found for earlier-type spirals.  The numbers
presented here are estimated in a manner similar to that described 
in $\S4.5.1$.}
\tablenotetext{b}{The technique used by Goudfrooij et~al.\  (2003) to 
estimate the total number of GCs assumes a fixed gaussian width, but
fits for the peak in the GC luminosity function.
For NGC~3628, the
peak luminosity appears to be more than 0.5 magnitudes fainter
than the expected turnover.  The value given here is 
based on an estimate where the peak is assumed to be located
at $M_V=-7.4$.}
\normalsize
\enddata             
\end{deluxetable}

\subsection{Is There a Case for ``Universal'' metal-rich bulge globular cluster systems?}

In the Galaxy,
the inner metal-rich GC subsystem has metallicity, kinematics,
and spatial distributions comparable to those of the underlying bulge
stars (e.g., Minniti 1995; Cote 1999).  
Forbes et~al.\  (2001) extended this evidence to suggest that the number
of red, metal-rich GCs normalized by the bulge luminosity is constant
in nearby spirals.  They compared the GC systems of three spiral
galaxies of different type (Milky Way=Sbc, M31=Sb, and M104=Sa), and
argued that the bulge specific frequency, which they define as the
number of metal-rich GCs within 2 bulge effective radii from the
galaxy center divided by the total bulge luminosity, was consistent
among these three galaxies, and similar to values found for field
ellipticals.  A similar pattern is seen in elliptical and lenticular
galaxies, where the spatial distribution of metal-rich GCs (typically)
closely follows that of the spheroidal light distribution, whereas the
(blue) metal-poor GCS is usually more extended (e.g., Kundu \&
Whitmore 1998; Puzia et~al.\  1999; Barmby et~al.\  2002).  If an
association between the bulges of spirals and their metal-rich GC
systems can be established, this would provide an important link
between the formation of spheroidal systems in general and the
formation of metal rich GCs.

An alternative scenario is the possibility that bulges form from the
redistribution of angular momentum of inner disk stars via bar
instabilities.  In this ``secular'' evolution (e.g., Pfenniger \&
Norman 1990), galaxies can evolve along the Hubble sequence from late-
to early- types.  However, since only disk stars contribute to bulge
formation in this model, GCs are not involved, and hence no
(metal-rich) bulge GCs are expected.

In their study of 7 edge-on disk systems, Goudfrooij et~al.\  (2003)
found evidence supporting an association of inner metal-rich GCs with
spiral bulges.
Their findings appear inconsistent with secular evolution.
However, one bulge-dominated spiral galaxy, NGC~7814, did not show
evidence for a system of inner, metal-rich GCs, implying that secular
evolution is still a viable option for some systems. 

In this section, we focus on the two earliest-type spirals in our
sample, M81 and M51, where the number of GCs associated with bulge
formation should be highest.  In $\S4.3$, we noted the apparent lack
of metal-rich GCs in M51.  Here, we quantify the expected number of
metal-rich GCs associated with the M51 bulge, and compare with
predictions based on the Forbes et~al.\  (2001) scenario.
Figures~\ref{2colmetal2}a shows that there are four
GCs in our M51 sample which appear to be metal-rich.
Assuming that we are missing the faint portion of the GC luminosity
function (correction factor $\sim2$) and that there are a similar
number of metal-rich GCs behind the disk, we estimate
that M51 has no more than 16 metal-rich GCs associated with the bulge.
Here we determine the number of metal-rich GCs predicted by the
``universal'' bulge system scenario.  Assuming a bulge $S_N$ of 0.5 as
given by Forbes et~al.\  (2001), and using the bulge $M_V$ value given
in Table~9, 98 red GCs are predicted to reside within 2 bulge
effective radii of M51.  Because bulge luminosities can be difficult
to estimate correctly from bulge/disk decompositions, we attempted to
use other techniques to estimate the M51 bulge luminosity as well.
The smallest estimate comes from the $\sigma-\mbox{black hole mass}$
relationship, and gives $M_V$ bulge $=-19.8$ (Wu \& Han 2001).  This
bulge luminosity combined with the Forbes model ($S_N=0.5$) predicts
45 red GCs.  This is still roughly a factor of $3$ larger than
implied by our observations.

Therefore, we tentatively conclude that M51 does not follow the
``universal'' bulge GC population as suggested by Forbes et
al. (2001).  
Because bar formation can be triggered via galaxy interactions, and
M51 shows the clear signs of on-going interaction with its nearby
barred companion NGC~5195 (one manifestation may be the many young
stars and clusters seen in the central region; Lamers et~al.\  2002), we
suggest that M51 is consistent with the idea of secular evolution
resulting in bulge formation.
An alternative explanation is that the $\sigma-\mbox{black hole mass}$
overestimates the total bulge luminosity by a factor of $\sim3$ in M51.

Approximately 40\% of our M81 GC sample appears to be metal-rich.
Based on the total number of GCs estimated for M81, we
calculate $S_N=1.0^{+0.4}_{-0.3}$ 
for the bulge normalized specific frequency, assuming the B/T ratio
given in Table~9 and that 40\% of the total M81 GC population is
metal-rich.  This range is similar to that found for the bulge
dominated systems NGC~3628 and NGC~4594 (Goudfrooij et~al.\  2003).

Most of the red clusters discovered in this work are located beyond
$2r_{eff}$ of the M81 bulge.  In fact, red clusters in our sample are
found in all studied WFPC2 fields, which cover a range of different
environments.  This is reminiscent of the metallicity/position
distribution of M31 GCs.  While we cannot directly comment on any
bulge GC population in M81 due to our inadequate coverage of the most
central portions, we note that Schroder et~al.\  (2002) found metal-rich
GCs residing beyond two bulge $r_{eff}$ (based on spectroscopy).
At face value, this suggests that M81 may also have retained a
system of rotating disk GCs.

We conclude that our study finds mixed support for the concept that
metal-rich GCs in spirals are associated with bulge formation.  M51 in
particular, which is similar in type to the Milky Way, appears to be
lacking a metal-rich bulge GC population, and metal-rich GCs in
general, despite some evidence that a central bulge does exist.  This
galaxy
is consistent with a scenario of bulge formation through secular
processes.

\subsection{Implications for Galaxy Formation}

Properties of GC systems in spiral galaxies are a promising tool for
understanding the formation histories of disks, bulges, and halos.  In
this work, we presented the photometric properties for GC systems in
five, nearby, low inclination spirals.  Future follow-up spectroscopy
for ages, abundances, and velocity information will provide additional
constraints on GC system properties, and the formation of spirals.  By
way of summarizing what is currently known about spiral GC systems, we
have compiled a ``scoresheet'' in Table~10.  The following properties
of GC systems are included in the indicated column: galaxy (1); total
V band luminosity (2); galaxy type (3); specific frequency, $S_N$ (4);
mass-normalized number of GCs, $T$ (5); magnitude of the GC luminosity
function turnover (upper luminosity limits are given in some cases)
(6); ratio of blue-to-red GCs (7); is the GC system consistent with a
``universal'' halo population? (8); did a constant (mass-normalized)
number of metal-rich GCs appear to have formed in association with the
bulge?  (9); is there a disk population? (10); is the size
distribution similar to that found in the Milky Way? (11); and finally
the references (12).

Here, we use the entire sample presented in Table~10 to try and
understand the formation of spiral GC systems in a broader context.
As mentioned earlier, Goudfrooij et~al.\  (2003) suggest that all
spirals, regardless of Hubble type, have formed a ``universal''
mass-normalized number of GCs in their halos.  They suggest that a
higher GC specific frequency should be observed in galaxies with
dominant bulges, since a second, metal-rich population is expected to
form associated with bulge stars.  Rhode (2003) however, found
evidence that the number of GCs depends primarily on the total
luminosity (mass) of the host galaxy, and has little to do with Hubble
type.  Since their conclusion is based on a sample including four
early-type galaxies and four spirals, 
it is not clear whether this is true for spirals in general.  
Below, we look for correlations between GC specific frequencies
and host galaxy properties.

\begin{figure}
\epsscale{0.6}
\plotone{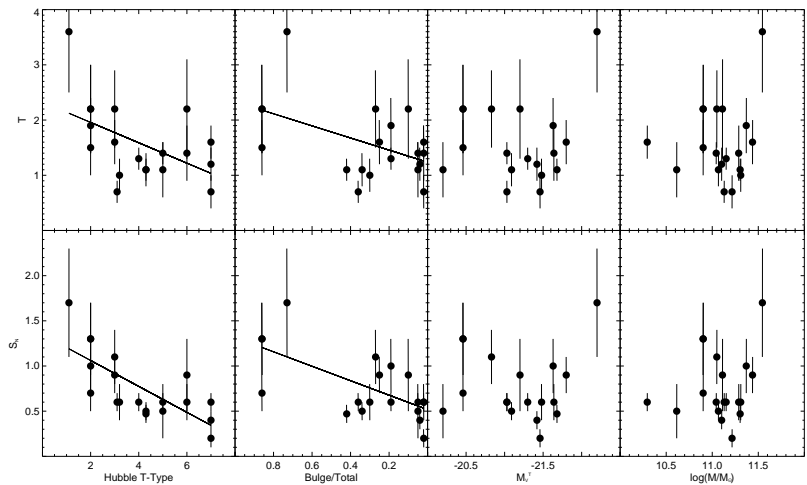}
\caption{Figures showing spiral GC systems' specific frequencies and T
values as a function of Hubble type, bulge/total, total V luminosity
of the galaxy, and galaxy mass.
\label{specfreq}}
\end{figure}

In Table~11 we compile mean and median values of
$S_N$ and T for spiral GC systems by dividing the total sample into
two groups: early vs. late-type, large vs. small bulge, bright
vs. faint, and high vs. low mass.  
These show that in general, the T parameter, which was devised to
roughly account for differences in the dominant stellar populations in
galaxies of different types, shows more scatter than the specific
frequency $S_N$.  To further quantify any trends, in
Figure~\ref{specfreq} we show $S_N$ and T as a function of galaxy
parameters (taken from Table~10).  A linear fit was performed for each
dataset, and is plotted if a correlation (at the $\gea 2\sigma$ level)
is found.  The slopes from the fits are also recorded in Table~11.
The strongest correlation $(4-5\sigma$) is between Hubble type and the
$S_N$ of GC systems, in the sense that earlier-type
spirals have larger numbers of luminosity normalized clusters.
The second column of plots also show a correlation between $S_N$ and
T with B/T ratio, although difficulties in bulge/disk
decompositions may contribute to the somewhat weaker correlation
($\sim3-4\sigma$) relative to that seen with Hubble type.
The last two sets of panels show no evidence for a correlation of GC
numbers with total luminosity or mass of the host galaxy.

\begin{deluxetable}{lllcccccr}             
\tablewidth{0pc}             
\tablecaption{MEAN SPECIFIC FREQUENCIES FOR GLOBULAR CLUSTER SYSTEMS IN SPIRALS} 
\scriptsize
\tablehead{             
\colhead{Group} &
\colhead{range} &
\multicolumn{3}{c}{$S_N$} &
\multicolumn{3}{c}{T} &
\colhead{N} 
\\
\colhead{} & 
\colhead{} & 
\colhead{median} &
\colhead{mean} & 
\colhead{slope\tablenotemark{a}} &
\colhead{median} &
\colhead{mean} & 
\colhead{slope\tablenotemark{a}} &
\colhead{} 
}             
\startdata           
1. early-type & Sa-Sbc & 0.90 & 0.90$\pm$0.11 & \nodata & 1.60 & 1.70$\pm$0.23 & \nodata & 12  \\
2. late-type & Sc-Scd & 0.60 & 0.54$\pm$0.10 & \llap{$-$}0.14$\pm$0.03 & 1.40 & 1.32$\pm$0.18 & \llap{$-$}0.18$\pm$0.08 & 7  \\
 & & &  &  &  &  \\
3. large bulge & $B/T > 0.2$ & 0.90 & 0.92$\pm$0.13 & \nodata & 1.60 & 1.72$\pm$0.27 & \nodata & 10  \\
4. small bulge & $B/T \le 0.2$ & 0.60 &0.60$\pm$0.08 & 0.80$\pm$0.23 & 1.40 & 1.42$\pm$0.15 & 1.10$\pm$0.49 &  9  \\
 & & &  &  &  &  \\
5. high mass & log($M/M_{\odot})$ $> 11.1$ & 0.60 & 0.72$\pm$0.12 & \nodata & 1.30 & 1.52$\pm$0.25 & \nodata & 11  \\
6. low mass  & log($M/M_{\odot})$ $\leq11.1$ & 0.70 & 0.83$\pm$0.12 & 0.25$\pm$0.30  & 1.60 & 1.66$\pm$0.17 & 0.38$\pm$0.57 & 8  \\
 & & &  &  &  &  \\
7. bright & $M_V \leq -21.25$ & 0.60 & 0.72$\pm$0.15 & \nodata & 1.30 & 1.53$\pm$0.28 & \nodata & 9 \\
8. faint & $M_V > -21.25$ & 0.70 & 0.81$\pm$0.10 & \llap{$-$}0.12$\pm$0.25  & 1.60 & 1.62$\pm$0.18 & \llap{$-$}0.05$\pm$0.13 & 10  \\
\tablecomments{Uncertainties in the mean are calculated as $\sigma/\sqrt{N}$
}
\tablenotetext{a}{Slopes from the best linear fit to spiral GC system
properties (Hubble type, bulge/total ratio, galaxy luminosity and mass)
as a function of $S_N$ and T are presented. }
\normalsize
\enddata             
\end{deluxetable}

\section{SUMMARY AND CONCLUSIONS}

We have studied the properties of GC systems in five nearby, low
inclination spirals using multifilter \textit{HST} WFPC2 imaging.  By using
morphological information, we are able to separate clusters from
individual stars, background galaxies, and blends.  Crude spectral
energy distributions from broadband filters or color information are
used to distinguish between young and ancient clusters, and we detect
GC systems in all five target galaxies.  We find that the U band is
crucial to separate reddened young clusters from ancient clusters.
Below, we summarize the main conclusions from our study.

1. Based on estimated intrinsic colors, the M81 GC system has an
extended metallicity distribution.  This argues
for the presence of both metal-rich and metal-poor GCs.
This extended nature was not obvious in a previous study, when only
single (BVI) colors were tested.  However, by combining two color
measurements together and including the U band, we find evidence that
the color distributions are similar to those in the Andromeda and
Milky Way GC systems.  The M101 GC sample also has an extended color
distribution.  By contrast, our M51 cluster sample has a narrower and
bluer color distribution, similar to that of blue, metal-poor Galactic
globulars.  We suggest that the lack of inner, red GCs in M51 is
consistent with a secular origin for the M51 bulge.

2. The GC luminosity distributions for M101 and possibly NGC~6946, two
of three later-type galaxies, appear to continue
increasing to magnitudes fainter than $M_V\sim-7.4$, the expected
turnover in the GC luminosity function.  This is similar to what we
found earlier in M33, another late-type spiral.  In general, the
colors of these excess, faint clusters do not differ significantly
from those of old, open clusters in the Galaxy, or from more luminous
GCs in the same galaxies.  We suggest that faint GC candidates in
later-type spirals may be either intermediate age ($3-9$ Gyr) disk
clusters, or possibly the low mass extension of the original GC
population, which survived destruction due to different dynamical
conditions in later-type spirals when compared with earlier type
galaxies.

3. We made a comparison of the effective radii distributions for GCs in
our five target galaxies.  All but M51 have distributions which are
similar to that in the Milky Way.  GCs in M51 however, are more
extended on average, and $\sim1/2$ our sample has sizes consistent
with the ``faint fuzzies'' discovered in NGC~1023, although the 
colors and luminosities are quite different.

4. We find that the total GC populations in later-type spirals, and
the blue subsystems in earlier-types are consistent with forming a
``universal'' halo population, as suggested by Goudfrooij et
al. (2003), with T$=1.3\pm0.2$ and $S_N=0.5\pm0.2$.

5. Finally, by combining our sample with specific frequency values for
other nearby spirals taken from the literature, we find that overall T
and $S_N$ increase with morphology, from late- to early-type spirals
(if faint excess clusters in later-type spirals are excluded from the
calculation).  This is consistent with a scenario where all massive
spiral galaxies form a relatively constant number of halo GCs, and
earlier types formed an additional, metal-rich bulge and/or disk
population.  We find no tendency for more luminous or massive spirals
to have larger normalized GC populations than their less
luminous/massive counterparts.

\acknowledgements We are grateful to H. Lamers for sending us his age
fitting code, and to P. Barmby for providing M31 cluster extinction
estimates.  We thank J. Gerssen for his help with bulge/disk
decompositions, and P. Goudfrooij and Dean McLaughlin for useful
discussions.  Finally, we thank the anonymous referee, whose
suggestions improved the presentation of this paper.  M.G.L. was
supported in part by the ABRL (R14-2002-058-01000-0) and the BK21
program. R.C. is grateful for support from NASA through grant
AR-09192.01-A from the Space Telescope Science Institute, which is
operated by the Association of Universities for Research in Astronomy,
Inc., for NASA under contract NAS5-26555.

\clearpage

\clearpage

\end{document}